\documentclass[acmsmall, screen, nonacm, table]{acmart}
\renewcommand\footnotetextcopyrightpermission[1]{}

\makeatletter
\let\@authorsaddresses\@empty
\makeatother

\usepackage{graphicx}
\usepackage{listings}
\usepackage{algorithm}
\usepackage{algorithmic}
\usepackage[frozencache,cachedir=minted-cache]{minted}
\usepackage{enumitem}
\usepackage{multirow}
\usepackage{xcolor}
\usepackage{subcaption}

\begin{document}

\title{NineToothed: A Triton-Based High-Level Domain-Specific Language for Machine Learning}

\author{Jiacheng Huang}
\affiliation{
  \institution{Qiyuan Lab}
  \country{China}}

\author{Zimin Li}
\affiliation{
  \institution{Qiyuan Lab}
  \country{China}}

\author{Yinghui Li}
\affiliation{
  \institution{Qiyuan Lab}
  \country{China}}

\author{Haojie Wang}
\affiliation{
  \institution{Tsinghua University}
  \country{China}}

\begin{abstract}

The emergence of deep learning domain-specific languages (DSLs) has substantially reduced the obstacles in developing high-performance, cross-platform compute kernels. However, current DSLs, such as Triton~\cite{10.1145/3315508.3329973}, still demand that developers possess expertise in parallel programming and expose them to many low-level details. This requirement complicates the development process and adds to the difficulty of maintaining compute kernels. Consequently, developing a new programming model that supports serial programming for deep learning workloads is crucial.

This paper introduces NineToothed, a domain-specific language that offers serial semantics for machine learning programming. Through the automatic transformation of serial code into parallel code, NineToothed significantly streamlines the development process while causing minimal performance degradation. NineToothed encompasses (1) a language with tensor-oriented metaprogramming (TOM) that adopts the arrange-and-apply paradigm, enabling the expression of tiled computations without the need to manage low-level details and (2) a code generator for generating high-performance parallel code. Our evaluation results indicate that NineToothed can greatly simplify compute kernel development while maintaining performance comparable to that of Triton.

\end{abstract}

\maketitle

\section{Introduction}

Driven by recent advancements in the artificial intelligence (AI) industry, the AI accelerator sector has increasingly diversified, with vendors developing their own hardware architectures and programming models, such as NVIDIA's CUDA and Cambricon's BANG. However, these architectures and models often differ significantly and are incompatible, raising the barrier for cross-platform AI software development. The emergence of domain-specific languages (DSLs) for deep learning, such as Triton~\cite{10.1145/3315508.3329973}, has helped mitigate this challenge. Triton, backed by the MLIR compiler infrastructure~\cite{lattner2020mlircompilerinfrastructureend}, can be transformed through multiple levels of intermediate representations (IRs) to generate executable high-performance compute kernels. This infrastructure allows vendors to add their compiler backends to generate machine code for their platforms. Thus, by developing compute kernels with Triton, developers can achieve cross-platform compatibility with minimal effort.

However, despite abstracting away many low-level details of programming models such as CUDA, Triton still exposes several underlying complexities. Developers are still required to perform pointer arithmetic and manage memory access explicitly to locate the relevant data for operation. This reliance on manual handling increases code volume, reduces readability, and complicates debugging and maintenance.

This paper proposes NineToothed, a domain-specific language that offers serial semantics for parallel programming. In the design process of NineToothed, three issues are primarily addressed. First, how can essential parallel information be integrated into serial code while maintaining code conciseness? Second, how can the transformation from serial code to parallel code be automated? Third, how can the performance of the generated parallel code be ensured?

To address the challenges of parallel information integration and code simplification, we design a tensor-oriented metaprogramming (TOM) technique. By introducing symbolic tensors that encapsulate symbolic information, the compiler can perform meta-operations, i.e., compile-time manipulations of tensor structures, which enables the formulation of tensor arrangements that embed parallel information. The applications of these arranged tensors can be defined independently and subsequently integrated with the tensor arrangements, which produces parallel compute kernels. This arrange-and-apply paradigm abstracts away low-level concerns such as pointer arithmetic and memory access, significantly simplifying NineToothed's code.

NineToothed's code generator tackles the problems of serial-to-parallel code transformation and performance optimization using two key mappings: tile-to-program mapping and source-to-target mapping. The tile-to-program mapping associates the serial code with each tile, while the source-to-target mapping associates each tile with its data. The mapping algorithms are carefully designed to ensure that the parallel workflow closely mirrors Triton's. Therefore, the generated parallel code achieves performance almost identical to those written directly in Triton.


We evaluate NineToothed on an NVIDIA A100 80GB PCIe GPU by implementing a set of common compute kernels using both NineToothed and Triton, then comparing their code and performance. Experimental results shows that the Halstead code volume of compute kernels implemented with NineToothed ranges from only 0.25\% to 56.33\% of that written using Triton, while the performance of the resulting kernels is comparable to that of Triton.

In summary, we make the following contributions:

\begin{itemize}
    \item We propose tensor-oriented metaprogramming, which enables a clear expression of tensor arrangements at compile time through meta-operations such as \texttt{tile}, \texttt{expand}, and \texttt{flatten}.
    \item We introduce an arrange-and-apply paradigm that divides an algorithm into arrangement and application of tensors, simplifying typical parallel programming models by abstracting away low-level concerns such as pointer arithmetic and memory access.
    \item We develop a code generator that can automatically transform serial code into parallel code through tile-to-program mapping and source-to-target mapping techniques.
    \item We implement NineToothed and integrate it into PyTorch. An extensive evaluation is conducted to compare NineToothed's code simplicity and performance with Triton's. The results show that the code written in NineToothed is significantly more concise and readable than that written in Triton while maintaining performance on par with equivalent code written in Triton.
\end{itemize}

This paper is organized as follows. Section~\ref{sec:background} introduces the background and motivation of NineToothed. Section~\ref{sec:design} explains the design details of NineToothed, including the tensor-oriented metaprogramming technique and the arrange-and-apply paradigm. Section~\ref{sec:kernel} demonstrates how NineToothed can be used to implement common compute kernels. Section~\ref{sec:evaluation} evaluates how well NineToothed simplifies the code while maintaining performance. Section~\ref{sec:related_work} introduces related work. Section~\ref{sec:conclusion} is the conclusion.

\section{Background and Motivation}
\label{sec:background}

An essential component of AI models is the operator, and the introduction of new model architectures often brings along new operators. For example, the transformer architecture relies fundamentally on the attention operator~\cite{NIPS2017_3f5ee243}, while the convolutional neural network (CNN) architecture is built upon the convolution operator~\cite{7780459}. Therefore, the ability to clearly and efficiently express and develop compute kernels, the core of operators, can facilitate the prototyping and validation of new operators and model architectures. However, developing compute kernels for different hardware platforms presents significant challenges. Each platform, such as NVIDIA's CUDA or Cambricon's BANG, has its own programming model and optimization strategies, which vary in memory management, parallelism, and instruction sets. This cross-platform compute kernel immigration problem is greatly mitigated by Triton, a DSL designed to simplify the development of high-performance compute kernels for modern AI workloads. By leveraging different dialects, vendors can develop their own compilation backends to support Triton. This fact implies that Triton can efficiently target multiple hardware architectures with the same high-level code, making it possible to develop cross-platform compute kernels without needing to rewrite code for different GPU architectures.

Although Triton has taken a significant step forward in developing high-performance, cross-platform compute kernels and has abstracted away many low-level details, it is still similar to other lower-level parallel programming models, e.g., CUDA and BANG, in requiring developers to use pointer arithmetic and explicit memory access. For instance, Triton provides the \texttt{program\_id} function, analogous to the built-in variables in CUDA, such as \texttt{threadIdx}, \texttt{blockDim}, and \texttt{blockIdx}, as well as those in BANG, like \texttt{taskId} and \texttt{clusterId}. For developers familiar with serial programming but new to parallel programming, such a programming model can be counterintuitive.

Consider the two snippets of pseudocode in Listings~\ref{listing:typical_serial_and_parallel_programming_models}. The snippet on the left represents a typical serial programming model, where the outer function uses techniques such as iteration to locate the object to manipulate, and the inner function focuses solely on the object and the associated manipulations. This approach of encapsulation is intuitive and maintains a clear separation of concerns. In contrast, the snippet on the right illustrates a typical parallel programming model, where the outer function only handles the launch of multiple instances of the inner function, and each instance must independently locate the object it needs to manipulate before performing the manipulation. This approach is somewhat counterintuitive and blends the logic of object positioning with the logic of object manipulation.

\begin{listing}[h]
\begin{minipage}{0.5\linewidth}
\begin{minted}{python}
# A typical serial programming model.
def inner(obj):
    ...

def outer(objs):
    for obj in objs:
        inner(obj)
\end{minted}
\end{minipage}
\begin{minipage}{0.49\linewidth}
\begin{minted}{python}
# A typical parallel programming model.
def inner(objs):
    obj = find_obj(objs, pid)
    ...

def outer(objs):
    inner[(dim,)](objs)
\end{minted}
\end{minipage}
\caption{Typical serial and parallel programming models.}
\label{listing:typical_serial_and_parallel_programming_models}
\end{listing}

Inspired by the discussion above, our goal is to design a programming model that enables developers to write serial code that can be automatically transformed into parallel code. As a result, we propose NineToothed, which employs tensor-oriented metaprogramming to achieve this goal.

\section{System Design}
\label{sec:design}

NineToothed composes two key designs: the serial language specification and the parallel code generation. We design a tensor-oriented metaprogramming technique that embeds parallel information through serial code. Then we introduce an arrange-and-apply paradigm to aid parallel code generation.

\subsection{Tensor-Oriented Metaprogramming}

Typical parallel programming models often retain low-level details, such as pointer arithmetic and memory access, to provide an explicit mechanism for expressing parallelism. Consequently, the challenge lies in abstracting away these low-level details, enabling developers to employ serial semantics for parallel programming. The key to achieving this abstraction is determining how parallel information can be effectively embedded within serial code. We identify the essential parallel information and demonstrate how it is embedded in NineToothed through tensor-oriented metaprogramming.

\subsubsection{Parallel Information Analysis}

By analyzing the compute kernels written in Triton, we can identify the following fundamental functions:

\begin{itemize}
    \item \texttt{triton.language.program\_id}
    \item \texttt{triton.language.arange}
    \item \texttt{triton.language.load}
    \item \texttt{triton.language.store}
\end{itemize}

\noindent
The general workflow for implementing compute kernels in Triton can be outlined as follows:

\begin{enumerate}
    \item Obtain the program ID(s) of the current program with \texttt{program\_id} and perform localization.
    \item Employ \texttt{arange} along with the program ID(s) to calculate offsets and masks of the data.
    \item Load the relevant data using \texttt{load}, leveraging the offsets and masks calculated.
    \item Perform the required computations.
    \item Store the results using \texttt{store}, again utilizing the precalculated offsets and masks.
\end{enumerate}

\noindent
To invoke a compute kernel, a \texttt{grid} function must also be defined, which specifies the number of programs to be launched.

Therefore, to abstract away low-level details, NineToothed needs to hide the functions mentioned above, as well as the workflow steps 1, 2, 3, and 5. Instead of explicitly writing runtime code with pointer arithmetic and memory access, the same parallel information can be expressed through compile-time serial code, which is clearer and more meaningful. To achieve this abstraction, NineToothed introduces symbolic tensors and meta-operations.

\subsubsection{Symbolic Tensors}

Inspired by Graphene~\cite{10.1145/3582016.3582018}, tensors in NineToothed are also designed to be hierarchical. Specifically, the data type of a tensor can itself be another tensor rather than being restricted to scalars such as floating-point or integer numbers. Additionally, certain tensor attributes, such as shape, can contain tuples rather than being limited to integers. Such a hierarchical tensor definition enables operations like tiling to be implemented and utilized more concisely. For example, a tensor $X_0$ of size $(4, 4)$ can be tiled into a tensor $X_1$ of size $(2, 2)$, where the data type of $X_1$ is a tensor $X_2$ of size $(2, 2)$.

This example involves concrete numerical values, but numerical tensor operations are insufficient for code generation because runtime data cannot be accessed at compile time. Observing a naive description of tiling in Algorithm~\ref{alg:tiling}, it is evident that while the number of dimensions of the tensor determines the number of iterations of the loop, other variables do not necessarily need to be numeric; they can also be symbolic.

\begin{algorithm}[h]
\caption{Naive tensor tiling that handles shapes only.}
\label{alg:tiling}
\begin{algorithmic}
\REQUIRE Tensor shape $\mathbf{shape}$ and tile shape $\mathbf{tile\_shape}$.
\ENSURE Shapes of tiled tensors.
\STATE Initialize empty lists $\mathbf{outer\_shape}$ and $\mathbf{inner\_shape}$.
\FOR{$size$, $tile\_size$ \textbf{in} ($\mathbf{shape}$, $\mathbf{tile\_shape}$)}
    \STATE $new\_size \gets \lceil size \div tile\_size \rceil$
    \STATE Append $new\_size$ to $\mathbf{outer\_shape}$.
    \STATE Append $tile\_size$ to $\mathbf{inner\_shape}$.
\ENDFOR
\RETURN $\mathbf{outer\_shape}$ and $\mathbf{inner\_shape}$
\end{algorithmic}
\end{algorithm}

Consequently, an approach similar to that used in computer algebra systems (CASs) like SymPy~\cite{10.7717/peerj-cs.103} can be adopted, which is storing symbolic expressions in tensor attributes such as shape and strides. In this way, operations on tensors are transformed into operations on symbolic expressions, effectively constructing symbolic expression trees, which are highly useful for code generation. Such a tensor composed of symbolic attributes is referred to as a symbolic tensor, which can be instantiated by passing its number of dimensions, as demonstrated in Listing~\ref{listing:demonstration_of_a_symbolic_tensor}.

\begin{listing}[h]
\begin{minted}{text}
>>> x = Tensor(2, name="x")
>>> x.shape
(x_size_0, x_size_1)
>>> x.strides
(x_stride_0, x_stride_1)
\end{minted}
\caption{Demonstration of a symbolic tensor.}
\label{listing:demonstration_of_a_symbolic_tensor}
\end{listing}

\begin{figure}[h]
\centering
\includegraphics[width=0.9\linewidth]{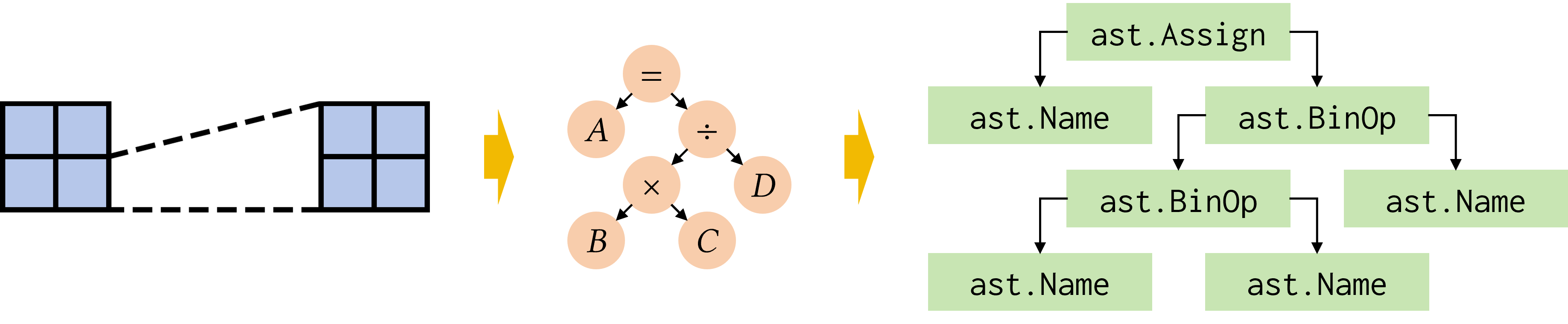}
\caption{In the AST-based implementation of symbolic tensors, tensor operations are essentially the construction of symbolic expression trees, which in turn are essentially the construction of abstract syntax trees.}
\label{fig:ast_based}
\end{figure}

Furthermore, observation shows that the symbolic expression trees involved in common tensor operations are often a subset of the abstract syntax trees (ASTs) of high-level programming languages. Since Triton currently utilizes Python's abstract syntax, rather than designing a new syntax and symbolic expression system for NineToothed from scratch and then converting the constructed expressions into Triton's AST, it is more efficient to leverage Python's syntax directly. By using Python's standard library \texttt{ast} as the foundation for symbolic expressions and wrapping the \texttt{ast.AST} class as expression tree nodes with appropriate implementation of magic methods, we can bring NineToothed closer to Triton and significantly improve the development efficiency of code transformation from NineToothed to Triton. Such an AST-based tensor implementation is shown in Figure~\ref{fig:ast_based}. This also allows NineToothed to be embedded within the Python ecosystem, making it easily callable by frameworks such as PyTorch or PaddlePaddle. Furthermore, since NineToothed generates Triton code, it naturally integrates into Triton's compilation flow, enabling seamless support across all hardware architectures and software platforms that Triton supports without any additional effort.

\subsubsection{Meta-Operations}

With the AST-based symbolic tensors that are optimized for code generation, we can now perform compile-time tensor operations, which are referred to as meta-operations.

\begin{table}[h]
\centering
\caption{NineToothed's meta-operations.}
\label{table:meta_operations}
\begin{tabular}{ll}
    \hline
    Operation & Description\\
    \hline
    \texttt{tile} & Forms a hierarchical tensor.\\
    \texttt{expand} & Expands singleton dimensions.\\
    \texttt{squeeze} & Removes singleton dimensions.\\
    \texttt{permute} & Permutes dimensions.\\
    \texttt{flatten} & Flattens dimensions.\\
    \texttt{ravel} & Flattens a hierarchical tensor.\\
    \hline
\end{tabular}
\end{table}

A list of meta-operations in NineToothed is shown in Table~\ref{table:meta_operations}, including the \texttt{tile} operation mentioned earlier. The \texttt{tile} operation is one of the most powerful features in NineToothed, offering more functionality than is immediately apparent in Algorithm~\ref{alg:tiling}. For example, \texttt{tile} includes several other parameters, such as the \texttt{strides} parameter, which controls the interval at which each tile is generated, analogous to the \texttt{stride} parameter in \texttt{torch.nn.functional.conv2d}. In fact, the \texttt{tile} operation shares similarities with a convolution, where the \texttt{tile\_shape} parameter corresponds to the filter shape. However, unlike the default stride of one in typical convolutional operations, the default value of \texttt{strides} in NineToothed is set to match the \texttt{tile\_shape}. This design choice is based on the observation that overlapping tiles are rarely required in many tiled computation algorithms, such as vector addition and matrix multiplication. While the \texttt{tile} operation includes other features that are beyond the scope of this discussion, we have focused on the \texttt{strides} parameter because it effectively illustrates how algorithms with complex tilings, such as convolution, can be efficiently implemented in NineToothed.

\begin{figure}[h]
    \centering
    \begin{subfigure}[c][1.5cm][c]{0.45\linewidth}
        \centering
        \includegraphics[scale=0.35]{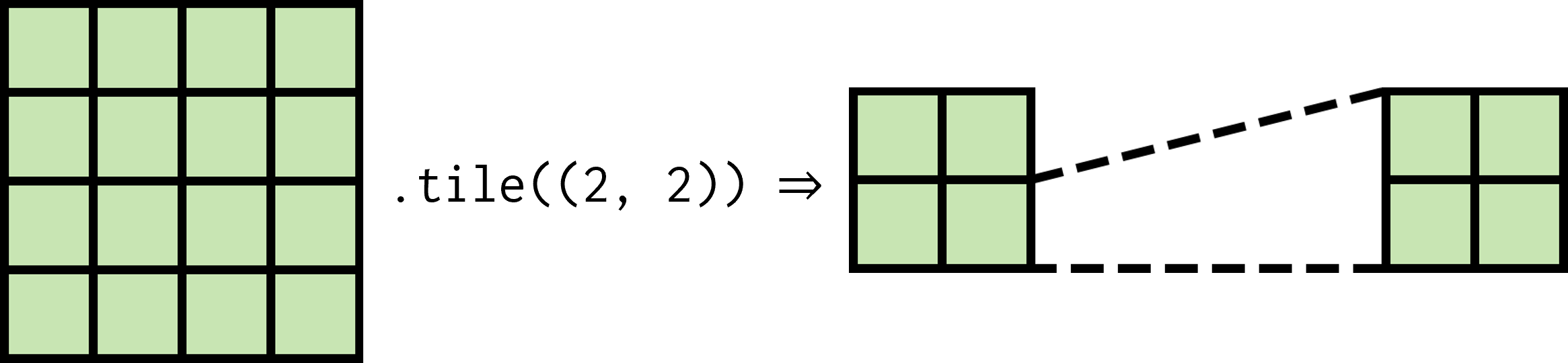}
    \end{subfigure}
    \begin{subfigure}[c][1.5cm][c]{0.45\linewidth}
        \centering
        \includegraphics[scale=0.35]{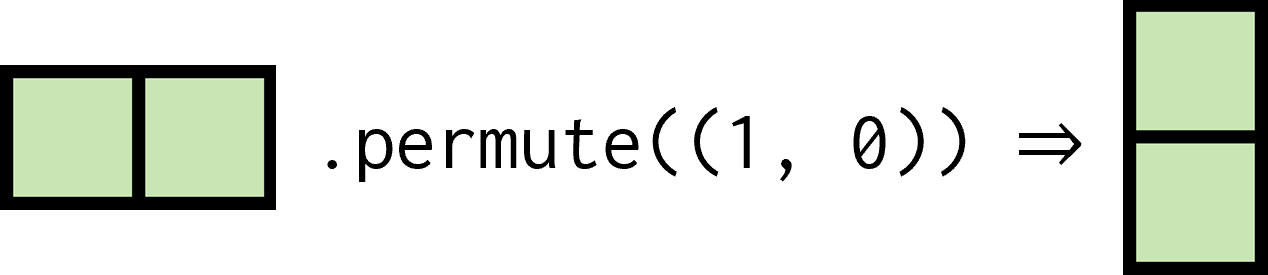}
    \end{subfigure}

    \begin{subfigure}[c][1.5cm]{0.45\linewidth}
        \centering
        \includegraphics[scale=0.35]{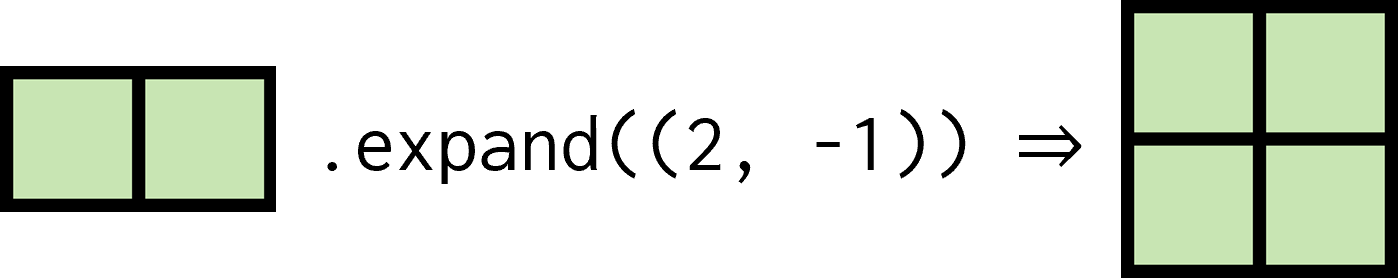}
    \end{subfigure}
    \begin{subfigure}[c][1.5cm]{0.45\linewidth}
        \centering
        \includegraphics[scale=0.35]{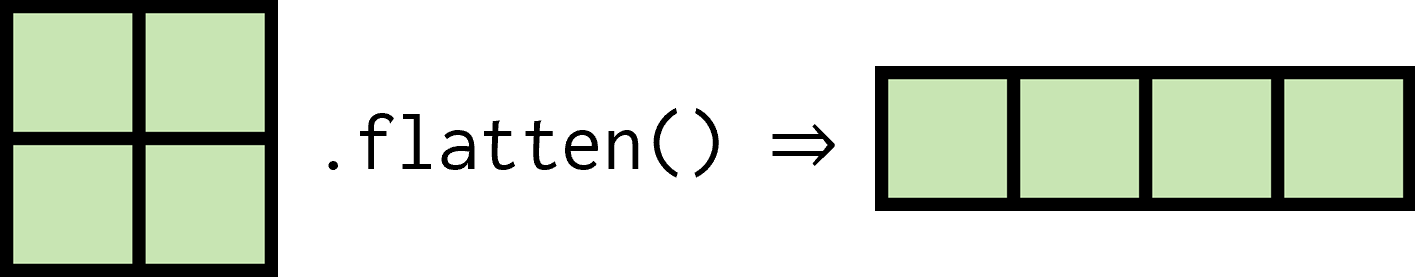}
    \end{subfigure}

    \begin{subfigure}[c][1.5cm]{0.45\linewidth}
        \centering
        \includegraphics[scale=0.35]{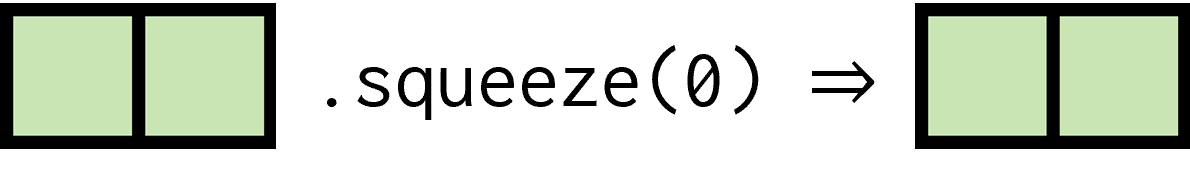}
    \end{subfigure}
    \begin{subfigure}[c][1.5cm]{0.45\linewidth}
        \centering
        \includegraphics[scale=0.35]{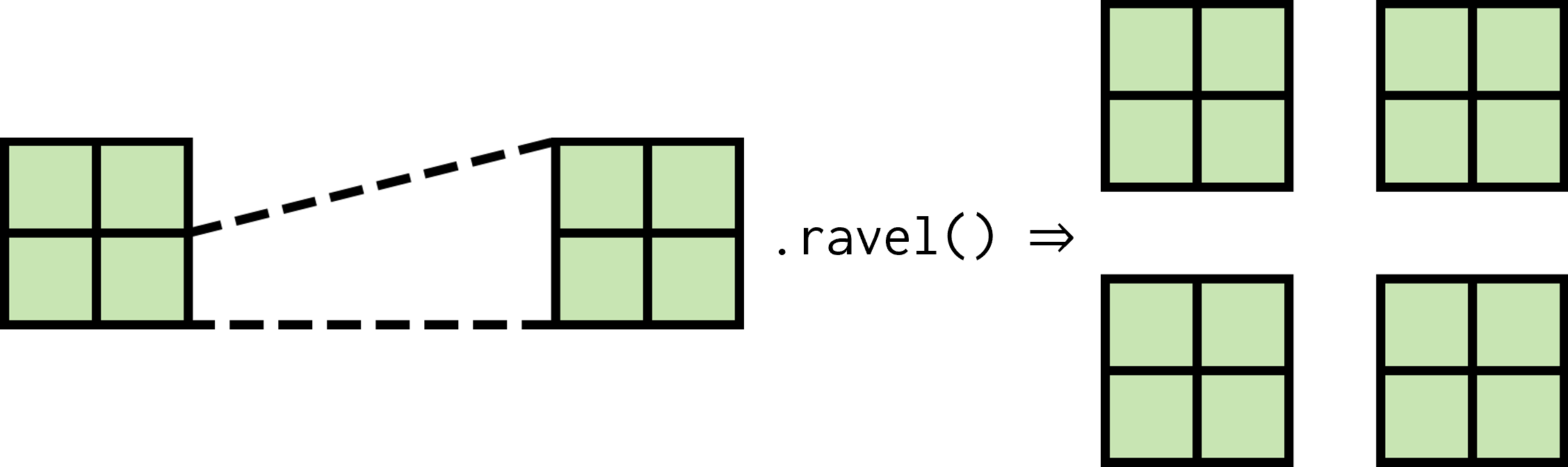}
    \end{subfigure}
\caption{A visualization of the meta-operations listed in Table~\ref{table:meta_operations}. The output tensor of the \texttt{squeeze} operation looks the same as the input tensor, but the input tensor has a shape of \texttt{(1, 2)}, while the output tensor has a shape of \texttt{(2,)}.}
\label{fig:meta_operations}
\end{figure}

Most of the other operations in NineToothed behave similarly to their counterparts in PyTorch. However, none of the meta-operations involves actual data movement, as no runtime data is available during compile time. The \texttt{ravel} operation, in contrast to its PyTorch namesake, behaves differently. Instead of flattening the specified dimensions of a single level of a hierarchical tensor into a single dimension like \texttt{flatten}, it flattens all levels of the tensor into a single level. For instance, consider a tensor with two levels: the first level has a shape of $(N, P, Q)$, and the second level has a shape of $(C, R, S)$. After applying \texttt{ravel}, the resulting tensor will have only one level with a shape of $(N, P, Q, C, R, S)$. A visualization of the meta-operations listed in Table~\ref{table:meta_operations} is provided in Figure~\ref{fig:meta_operations}.

\subsection{Arrange-and-Apply Paradigm}

\begin{figure}[h]
\centering
\includegraphics[width=0.9\linewidth]{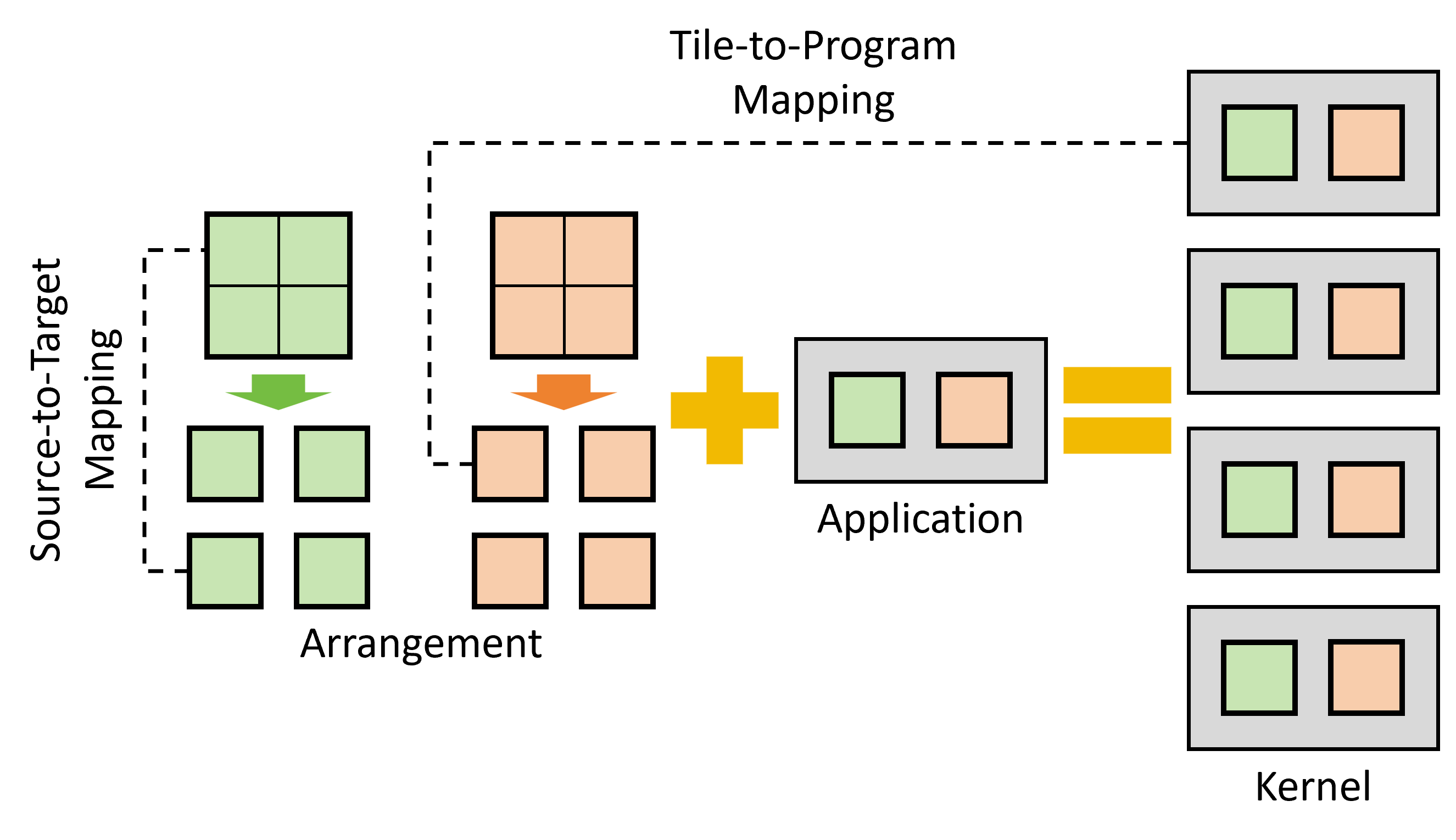}
\caption{An illustration of the arrange-and-apply paradigm.}
\label{fig:arrange_and_apply_paradigm}
\end{figure}

So far, we have described how to perform meta-operations on individual tensors, but there remains a gap in implementing a complete algorithm. In NineToothed, an algorithm is implemented following the arrange-and-apply paradigm that divides an algorithm into two parts: arrangement and application. In this paradigm, arrangement refers to performing meta-operations on tensors to align the tiles, while application refers to using the arranged tiles to carry out runtime computations. An illustration of this paradigm is shown in Figure~\ref{fig:arrange_and_apply_paradigm}.

This paradigm offers several advantages. First, separating arrangement and application allows for greater modularity, enabling the reuse of either component across algorithms with varying objectives. This separation facilitates more efficient development and reduces redundancy in compute kernel implementation. Second, the paradigm simplifies the analysis of tiling patterns. When arrangement and application are treated as a single unit within a compute kernel, it becomes challenging to determine whether different compute kernels share compatible tiling configurations, which is crucial for enabling compute kernel fusion. By isolating them, the paradigm promotes a clearer understanding of tiling compatibility and enhances opportunities for optimization.

\subsubsection{Tile-to-Program Mapping}

The first step of using the arrange-and-apply paradigm is to define the arrangement. As previously discussed, meta-operations in NineToothed work with hierarchical tensors, which are tensors composed of multiple levels. To ensure correct execution, a proper arrangement of an algorithm requires that the shapes of the outermost levels of the arranged parameter tensors be consistent. Any arrangement that results in mismatched shapes at the outermost levels signals an error. Building on this principle, NineToothed launches programs based on the shape of the outermost tensors of the arranged parameter tensors and maps the second outermost tensors to these programs. This approach allows each group of tiles at the outermost levels of the arranged tensors to correspond to a program. If there are more than two levels, the remaining levels can be accessed using the \texttt{[...]} syntax, similar to how elements are accessed in a container in Python. This is how the code generator establishes the tile-to-program mappings, as illustrated in Figure~\ref{fig:matrix_multiplication_arrangement}.

Since NineToothed follows the principle that the shapes of the outermost levels of the arranged parameter tensors are consistent, the \texttt{grid} function can be automatically generated. This is achieved by using the shape symbols from the outermost levels of one of the arranged parameters, with the number of programs calculated as the product of the sizes of all dimensions.

Furthermore, all other necessary information can be automatically obtained as long as a frontend of the target framework is available. For instance, in PyTorch, the shape and strides of a tensor can be accessed via \texttt{torch.Tensor.size} and \texttt{torch.Tensor.stride}, respectively. As a result, users are not required to provide this information manually. Instead, a launch function can be generated alongside the compute kernel, streamlining the process and reducing user burden.

\subsubsection{Source-to-Target Mapping}

Once the above mappings are established, the second step is to define the application, which specifies the runtime operations to be performed on each group of tiles. It is important to note that these operations are performed at the tile level, not at the tensor level. For example, a transpose operation is applied to each tile individually, rather than transposing the entire tensor directly.

Each group of tiles can be assigned to its corresponding program using the established tile-to-program mappings and the generated launch function. However, because parameters may undergo complex meta-operations during the arrangement phase, additional information is required to map data from the source tensors to the target tiles. To address this, each tensor in NineToothed stores two tuples: one representing the source dimensions and the other representing the target dimensions. The source dimensions specify the origins of the current tile's dimensions, while the target dimensions indicate their destinations. In terms of offset calculation, the source dimensions guide the code generator in determining which dimensions to reference when retrieving shape and stride information. Meanwhile, the target dimensions direct the code generator on how to broadcast the 1D offsets, constructed using \texttt{program\_id} and \texttt{arange}, to the appropriate dimensions. The broadcast offsets are then integrated into multidimensional offsets, which are used to locate data during the generation of \texttt{load} and \texttt{store} operations.

\subsubsection{Integration}

Once the arrangement and the application of the tensors are defined, the compute kernel can be generated by integrating them. As a result, the parallel programming functions previously mentioned, including \texttt{program\_id}, \texttt{arange}, \texttt{load}, and \texttt{store}, along with workflow steps 1, 2, 3, and 5, can be effectively abstracted away. Notably, it is not necessary to abstract away step 4, as it is inherently a serial programming step.

\section{Compute Kernel Implementations}
\label{sec:kernel}

In this section, several common compute kernels are implemented using NineToothed to demonstrate NineToothed's programming model.

\subsection{Vector Addition}

\begin{listing}[h]
\begin{minted}{python}
BLOCK_SIZE = Symbol("BLOCK_SIZE", constexpr=True)


def arrangement(input, other, output, BLOCK_SIZE=BLOCK_SIZE):
    input_arranged = input.tile((BLOCK_SIZE,))
    other_arranged = other.tile((BLOCK_SIZE,))
    output_arranged = output.tile((BLOCK_SIZE,))

    return input_arranged, other_arranged, output_arranged


def application(input, other, output):
    output = input + other


tensors = tuple(Tensor(1) for _ in range(3))

kernel = ninetoothed.make(arrangement, application, tensors)
\end{minted}
\caption{NineToothed implementation of vector addition.}
\label{listing:vector_addition}
\end{listing}

Listing~\ref{listing:vector_addition} presents an implementation of vector addition, a straightforward example commonly found in modern parallel programming models. While vector addition is relatively simple and may not fully showcase the advantages of NineToothed, it serves as an ideal introduction to understanding its programming model. This compute kernel performs the addition of \texttt{input} and \texttt{other}, storing the result in \texttt{output}.

Following the arrange-and-apply paradigm, this code consists of three main components, \texttt{tensors}, arrangement, and application, which are integrated via the \texttt{make} function to generate the \texttt{kernel}.

The expression \texttt{Tensor(1)} constructs a 1D tensor, or vector, so \texttt{tensors} is a tuple composed of three vectors, which are tiled into blocks of size \texttt{BLOCK\_SIZE} by the \texttt{tile((BLOCK\_SIZE,))} calls in the arrangement function. The default value for \texttt{BLOCK\_SIZE} is a \texttt{Symbol}, assigning it a name, and \texttt{constexpr=True} informs the compiler that \texttt{BLOCK\_SIZE} is a constant expression, meaning its value should be known at compile time.

In the application function, it is important to note that \texttt{input}, \texttt{other}, and \texttt{output} refer to individual blocks, not the entire tensors, so \texttt{output = input + other} performs element-wise addition on each group of blocks of \texttt{input} and \texttt{other}, storing the result in \texttt{output}. Since this operation is applied to each group of blocks, the addition is complete across all the tensors.

This \texttt{kernel} can be invoked as shown in Listing~\ref{listing:vector_addition_invocation}.

\begin{listing}[h]
\begin{minted}{python}
def add(input, other):
    output = torch.empty_like(input)

    kernel(input, other, output, BLOCK_SIZE=1024)

    return output
\end{minted}
\caption{Vector addition invocation.}
\label{listing:vector_addition_invocation}
\end{listing}

\subsection{Matrix Multiplication}

\begin{listing}[h]
\begin{minted}{python}
def arrangement(
    input,
    other,
    output,
    BLOCK_SIZE_M=block_size(),
    BLOCK_SIZE_N=block_size(),
    BLOCK_SIZE_K=block_size(),
):
    output_arranged = output.tile((BLOCK_SIZE_M, BLOCK_SIZE_N))

    input_arranged = input.tile((BLOCK_SIZE_M, BLOCK_SIZE_K))
    input_arranged = input_arranged.tile((1, -1))
    input_arranged = input_arranged.expand((-1, output_arranged.shape[1]))
    input_arranged.dtype = input_arranged.dtype.squeeze(0)

    other_arranged = other.tile((BLOCK_SIZE_K, BLOCK_SIZE_N))
    other_arranged = other_arranged.tile((-1, 1))
    other_arranged = other_arranged.expand((output_arranged.shape[0], -1))
    other_arranged.dtype = other_arranged.dtype.squeeze(1)

    return input_arranged, other_arranged, output_arranged
\end{minted}
\caption{The matrix multiplication arrangement function.}
\label{listing:matrix_multiplication_arrangement}
\end{listing}

\begin{figure}[h]
    \centering
    \begin{subfigure}[c][2cm][c]{0.3\linewidth}
        \centering
        \includegraphics[scale=0.09]{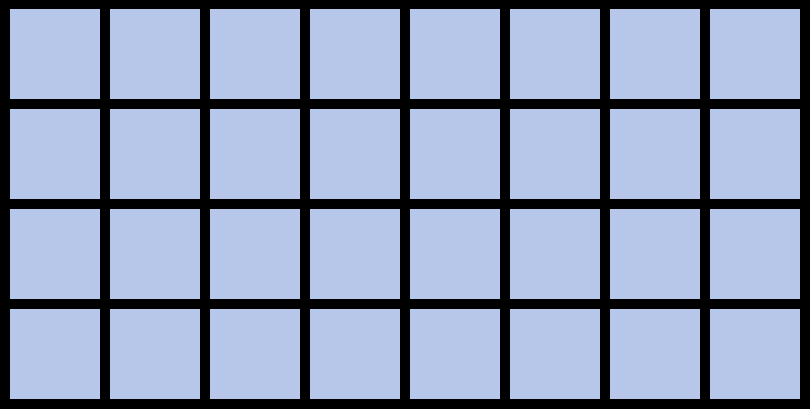}
    \end{subfigure}
    \begin{subfigure}[c][2cm][c]{0.3\linewidth}
        \centering
        \includegraphics[scale=0.09]{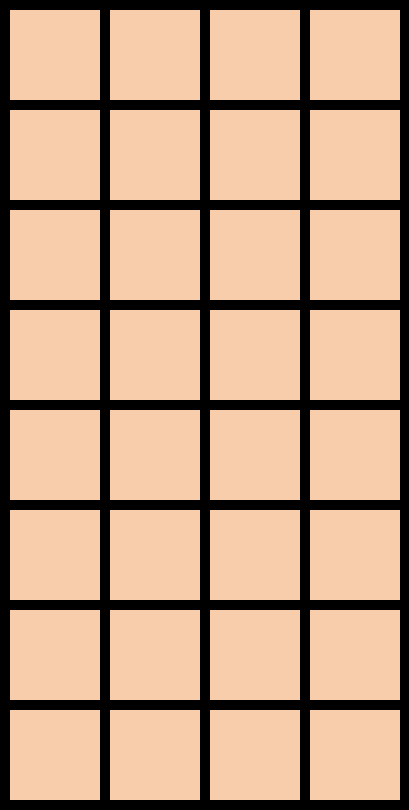}
    \end{subfigure}
    \begin{subfigure}[c][2cm][c]{0.3\linewidth}
        \centering
        \includegraphics[scale=0.09]{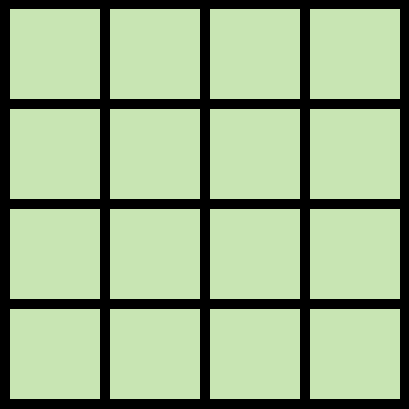}
    \end{subfigure}

    \begin{subfigure}[c][2cm][c]{0.3\linewidth}
        \centering
        \includegraphics[scale=0.09]{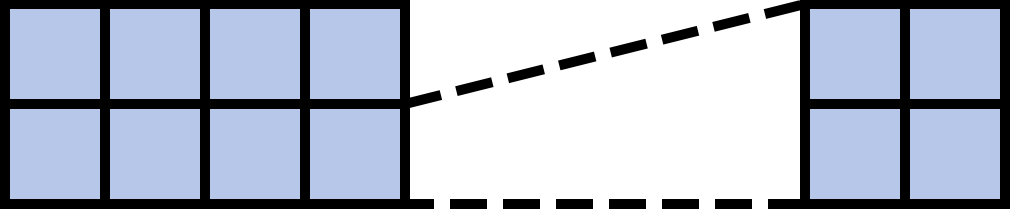}
    \end{subfigure}
    \begin{subfigure}[c][2cm][c]{0.3\linewidth}
        \centering
        \includegraphics[scale=0.09]{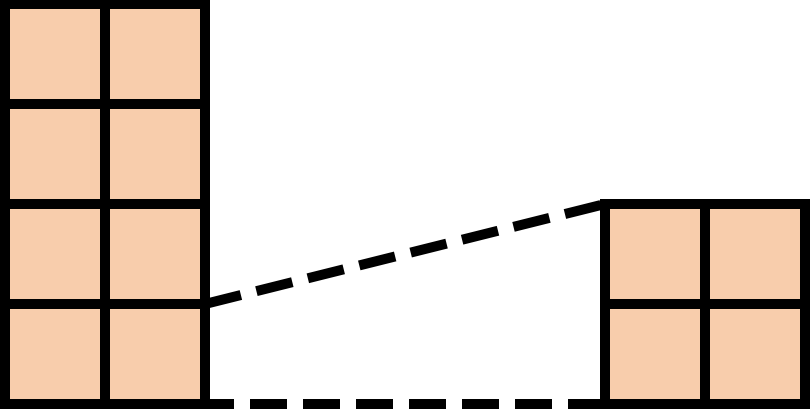}
    \end{subfigure}
    \begin{subfigure}[c][2cm][c]{0.3\linewidth}
        \centering
        \includegraphics[scale=0.09]{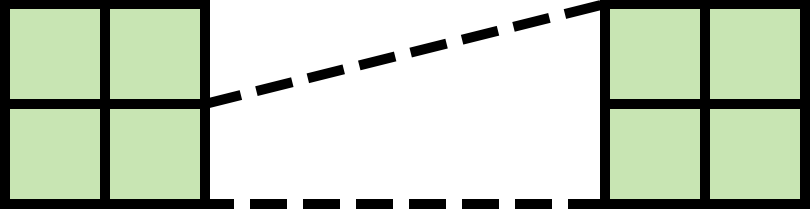}
    \end{subfigure}

    \begin{subfigure}[c][2cm][c]{0.3\linewidth}
        \centering
        \includegraphics[scale=0.09]{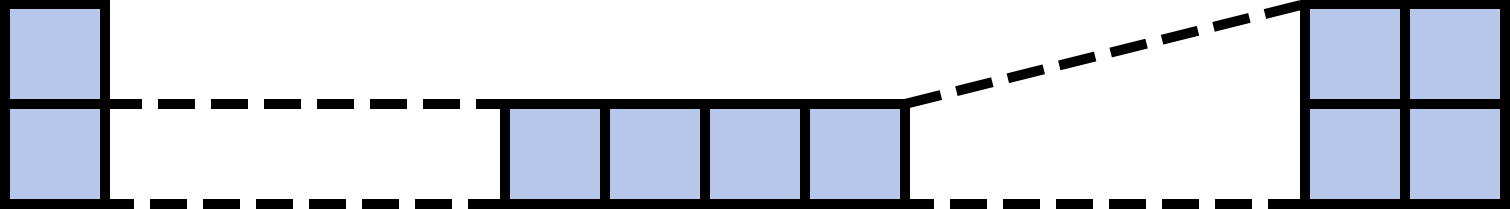}
    \end{subfigure}
    \begin{subfigure}[c][2cm][c]{0.3\linewidth}
        \centering
        \includegraphics[scale=0.09]{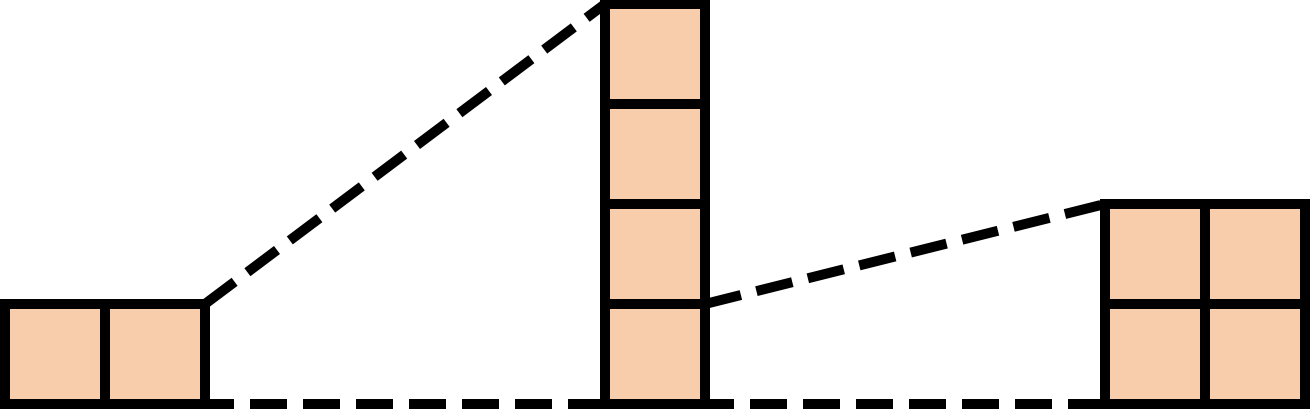}
    \end{subfigure}
    \begin{subfigure}[c][2cm][c]{0.3\linewidth}
        \centering
        \hfill
    \end{subfigure}

    \begin{subfigure}[c][2cm][c]{0.3\linewidth}
        \centering
        \includegraphics[scale=0.09]{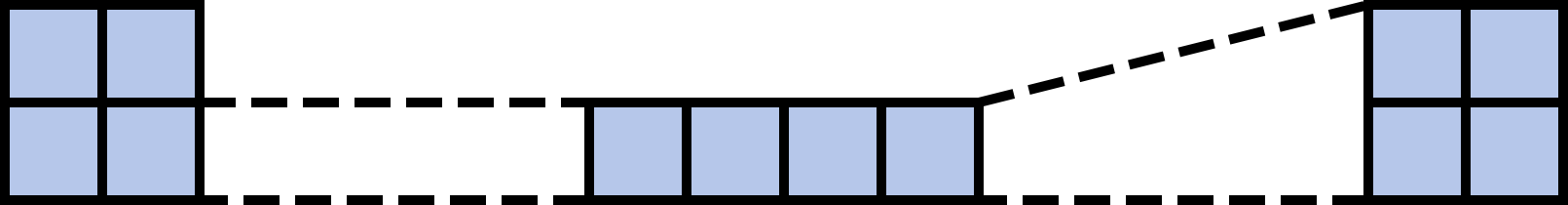}
    \end{subfigure}
    \begin{subfigure}[c][2cm][c]{0.3\linewidth}
        \centering
        \includegraphics[scale=0.09]{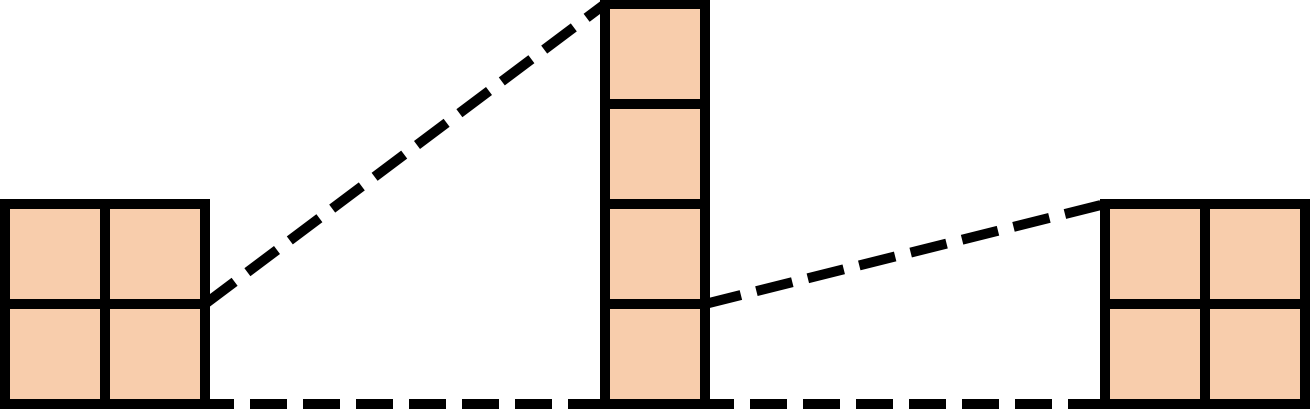}
    \end{subfigure}
    \begin{subfigure}[c][2cm][c]{0.3\linewidth}
        \centering
        \hfill
    \end{subfigure}

    \begin{subfigure}[c][2cm][c]{0.3\linewidth}
        \centering
        \includegraphics[scale=0.09]{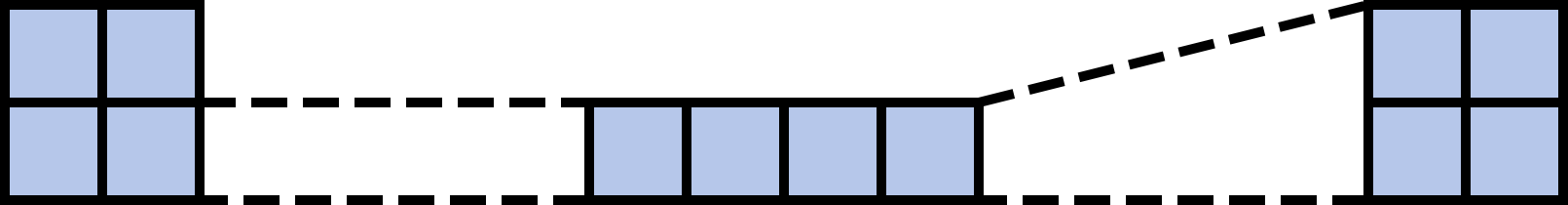}
    \end{subfigure}
    \begin{subfigure}[c][2cm][c]{0.3\linewidth}
        \centering
        \includegraphics[scale=0.09]{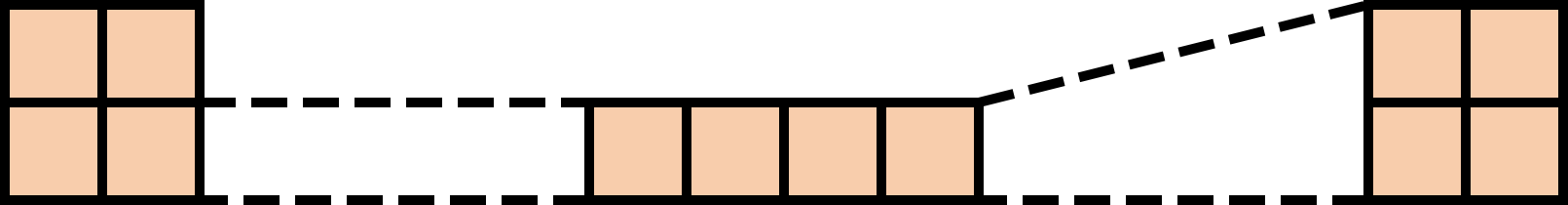}
    \end{subfigure}
    \begin{subfigure}[c][2cm][c]{0.3\linewidth}
        \centering
        \hfill
    \end{subfigure}
\caption{A visualization of the matrix multiplication arrangement. Although the source tensors have different shapes, a proper arrangement ensures that the shapes of the outermost levels of the arranged tensors are equal. Since these shapes are all \texttt{(2, 2)}, NineToothed will launch \texttt{4} programs, and the tiles of the outermost levels of the arranged tensors will be aligned to the programs launched.}
\label{fig:matrix_multiplication_arrangement}
\end{figure}

We now proceed to implement matrix multiplication. In this case, the arrangement function exhibits increased complexity, as illustrated in Listing~\ref{listing:matrix_multiplication_arrangement}. In this code, we have three tensor parameters: \texttt{input}, \texttt{other}, and \texttt{output}. Let us denote the matrices as $A$, $B$, and $C$, where $A$ and $B$ are the input matrices and $C$ is the output matrix.

Arranging $C$ is straightforward; it is tiled into blocks of size \texttt{(BLOCK\_SIZE\_M, BLOCK\_SIZE\_N)} along its rows and columns. However, arranging $A$ and $B$ requires additional steps. We introduce a meta-parameter \texttt{BLOCK\_SIZE\_K} to tile $A$ into blocks of size \texttt{(BLOCK\_SIZE\_M, BLOCK\_SIZE\_K)} and $B$ into blocks of size \texttt{(BLOCK\_SIZE\_K, BLOCK\_SIZE\_N)}. Unlike $C$, which maps directly block by block, the blocks of $A$ and $B$ must be aligned differently. Specifically, each row of $A$ must correspond to each column of $B$. To achieve this, we further tile $A$ by rows and $B$ by columns, resulting in sets of row blocks of $A$ and column blocks of $B$. These row and column blocks must be aligned, with each row block of $A$ corresponding to every column block of $B$. This alignment is achieved through the \texttt{expand} operation. We expand the row blocks of $A$ along the column dimension to match the number of columns in $C$ and expand the column blocks of $B$ along the row dimension to match the number of rows in $C$.

At this point, the arrangement can be considered complete, but we notice that the row and column blocks, which reside in two dimensions, have shapes of \texttt{(1, ...)} and \texttt{(..., 1)}, respectively. This results in access patterns like \texttt{input[0, k]} and \texttt{other[k, 0]}. If we want to use \texttt{input} to determine the range of \texttt{k}, we would need to refer to \texttt{input.shape[1]}. However, it is clear that singleton dimensions can actually be removed. To address this, we apply the \texttt{squeeze} operation to eliminate these singleton dimensions. This allows us to calculate the range via \texttt{input.shape[0]} and simplifies the access patterns to \texttt{input[k]} and \texttt{other[k]}. This approach successfully arranges $A$, $B$, and $C$, adhering to the principle that a proper arrangement ensures the shapes of the outermost levels of the arranged tensors are equal. An illustration of this arrangement can be found in Figure~\ref{fig:matrix_multiplication_arrangement}.

Next, let's proceed to the application function, as shown in Listing~\ref{listing:matrix_multiplication_application}. This function is straightforward: by iterating over the corresponding row blocks of $A$ and column blocks of $B$, we multiply them and accumulate the results in an accumulator. Finally, the accumulated result is stored in the corresponding block of $C$. Since this process is applied to each block, matrix multiplication is performed across all the tensors.

\begin{listing}[h]
\begin{minted}{python}
def application(input, other, output):
    accumulator = ntl.zeros(output.shape, dtype=ntl.float32)

    for k in range(input.shape[0]):
        accumulator += ntl.dot(input[k], other[k])

    output = accumulator
\end{minted}
\caption{The matrix multiplication application function.}
\label{listing:matrix_multiplication_application}
\end{listing}

Finally, with the arrangement and application functions defined, we can use the \texttt{make} function to integrate them into the matrix multiplication compute kernel, as shown in Listing~\ref{listing:matrix_multiplication_integration}.

\begin{listing}[h]
\begin{minted}{python}
tensors = (Tensor(2), Tensor(2), Tensor(2))

kernel = ninetoothed.make(arrangement, application, tensors)
\end{minted}
\caption{The integration of the matrix multiplication arrangement and application functions.}
\label{listing:matrix_multiplication_integration}
\end{listing}

\subsection{2D Convolution}

\begin{listing}[h]
\begin{minted}{python}
def arrangement(input, filter, output):
    input_arranged = input.tile((1, *filter.shape[1:]), strides=(-1, -1, 1, 1))
    input_arranged = input_arranged.squeeze(1)
    input_arranged.dtype = input_arranged.dtype.squeeze(0)
    input_arranged = input_arranged.ravel()
    input_arranged = input_arranged.flatten(end_dim=3).flatten(start_dim=1)

    filter_arranged = filter.flatten(start_dim=1)
    filter_arranged = filter_arranged.permute((1, 0))

    output_arranged = output.permute((0, 2, 3, 1)).flatten(end_dim=3)

    return mm.arrangement(input_arranged, filter_arranged, output_arranged)


shape_options = {"constexpr": True}
tensors = tuple(Tensor(4, shape_options=shape_options) for _ in range(3))

kernel = ninetoothed.make(arrangement, mm.application, tensors)
\end{minted}
\caption{NineToothed implementation of 2D convolution.}
\label{listing:2d_convolution}
\end{listing}

The previous example of matrix multiplication highlights the streamlined workflow enabled by \texttt{make}. However, the more significant advantage of \texttt{make} lies in its ability to leverage previously defined arrangement and application functions, thus eliminating the need to reinvent the wheel for each new task. Listing~\ref{listing:2d_convolution} demonstrates an implementation of 2D convolution based on the implicit GEMM algorithm~\cite{Thakkar_CUTLASS_2023}, which effectively maps convolution to matrix multiplication. Consequently, with \texttt{make}, we can reuse the arrangement and application functions already established for matrix multiplication. The only remaining step is to prepare the appropriate mappings from convolution to matrix multiplication and pass them to the matrix multiplication arrangement function. Once this is done, the rest of the process is automatically handled. As a result, no separate application function is required, and the algorithm can be implemented in NineToothed with minimal code.

\begin{figure}[h]
    \centering
    \begin{subfigure}[c][2.5cm][c]{0.3\linewidth}
        \centering
        \includegraphics[scale=0.04]{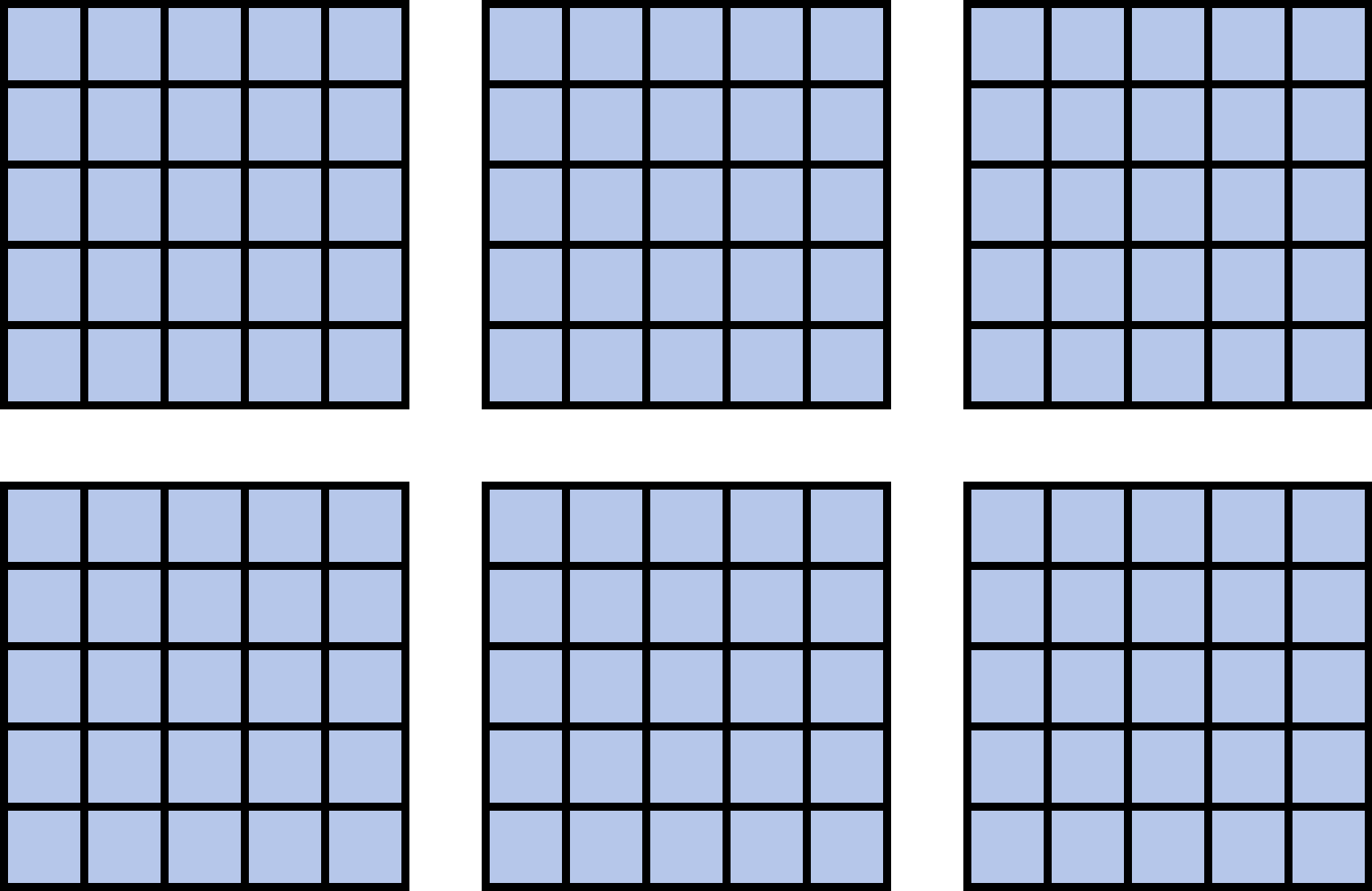}
    \end{subfigure}
    \begin{subfigure}[c][2.5cm][c]{0.3\linewidth}
        \centering
        \includegraphics[scale=0.04]{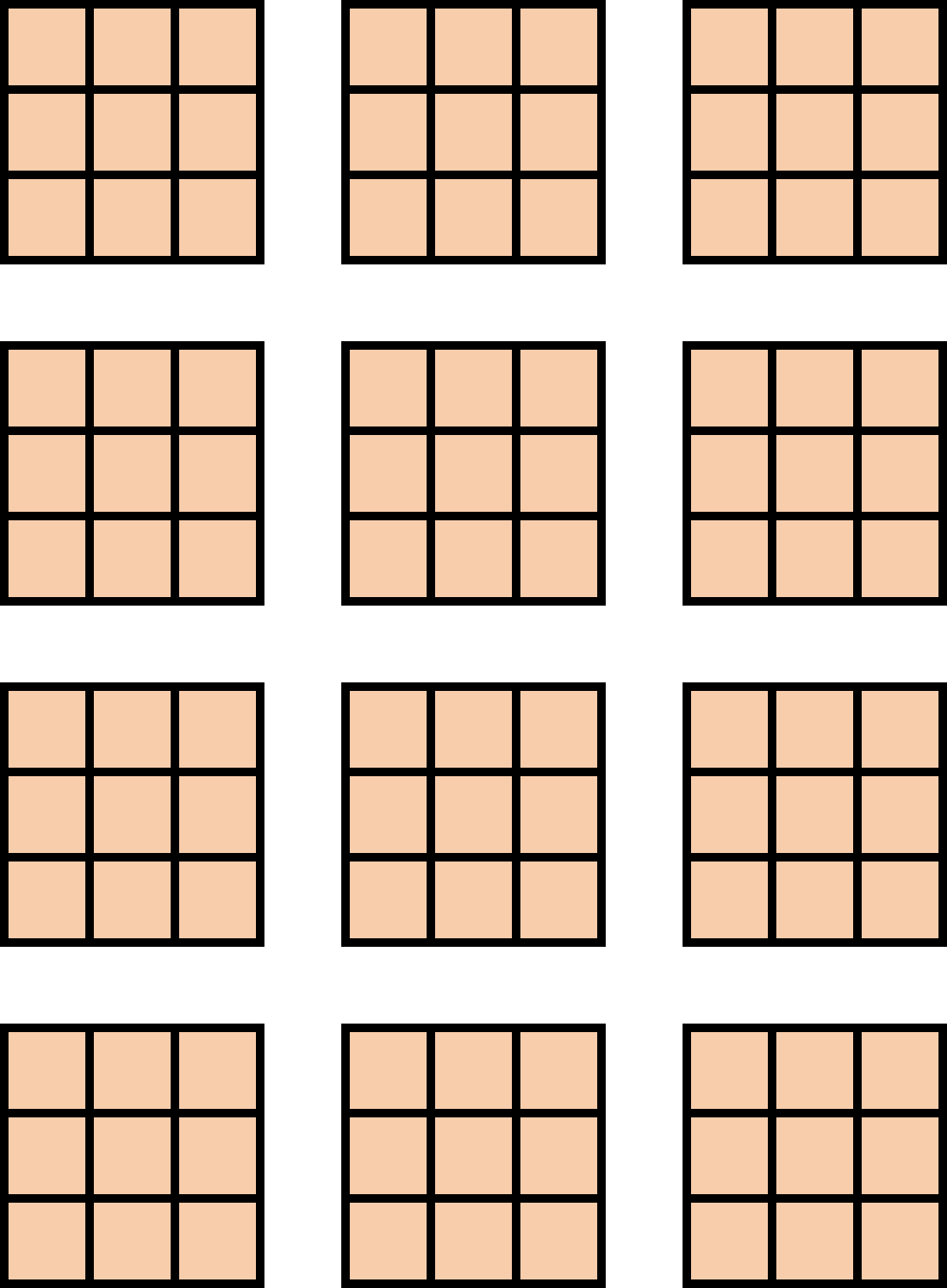}
    \end{subfigure}
    \begin{subfigure}[c][2.5cm][c]{0.3\linewidth}
        \centering
        \includegraphics[scale=0.04]{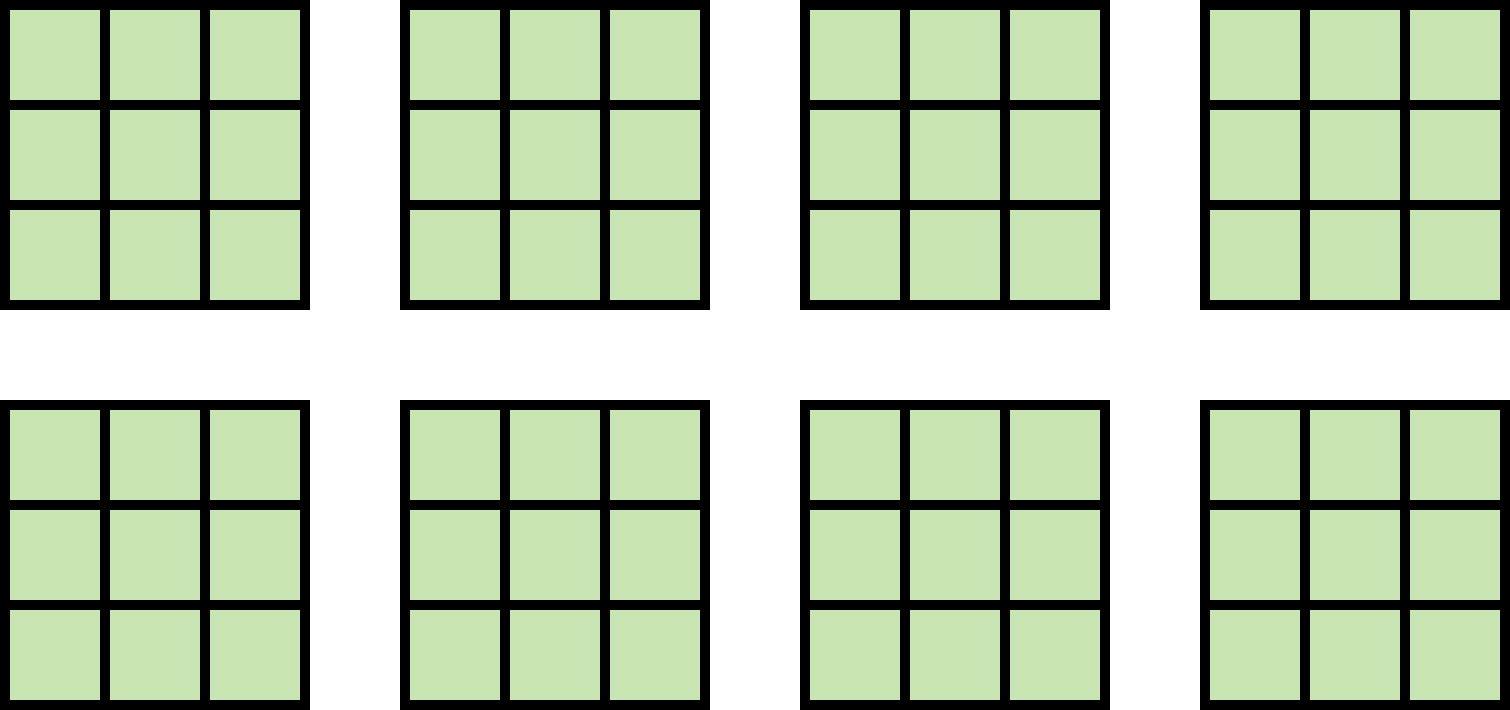}
    \end{subfigure}

    \begin{subfigure}[c][2.5cm][c]{0.3\linewidth}
        \centering
        \includegraphics[scale=0.04]{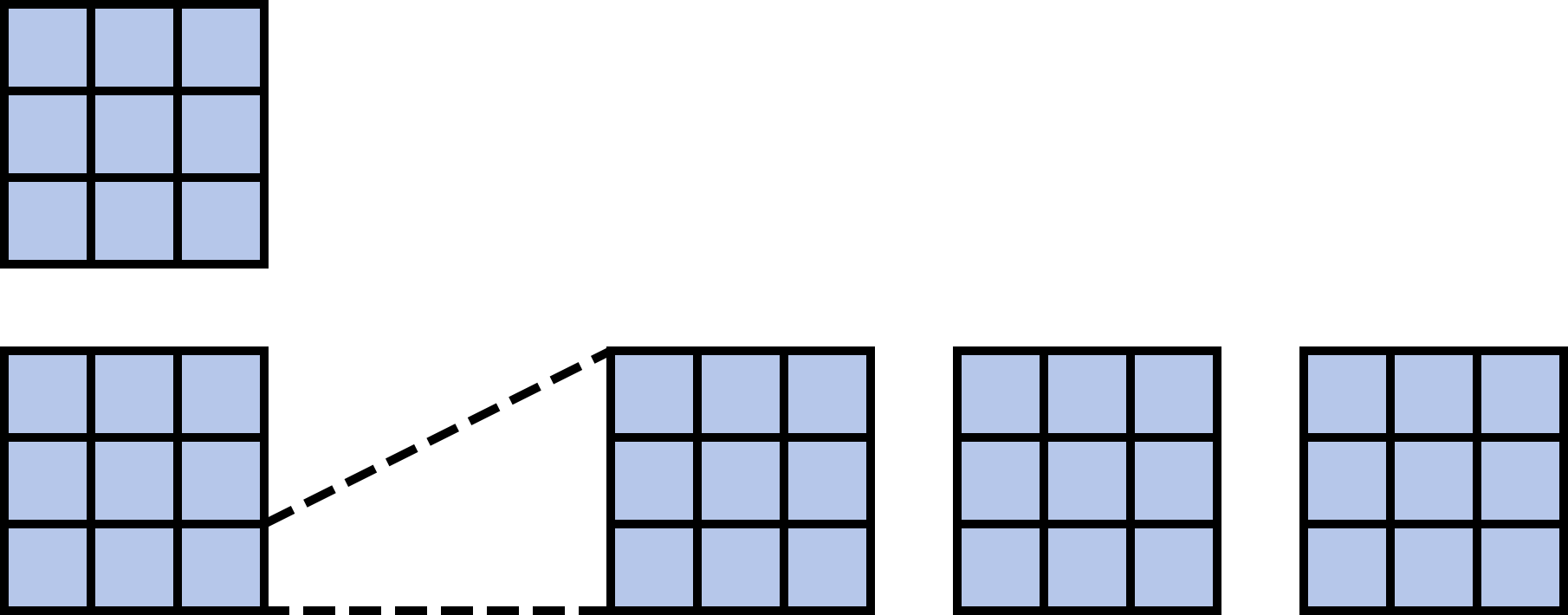}
    \end{subfigure}
    \begin{subfigure}[c][2.5cm][c]{0.3\linewidth}
        \centering
        \includegraphics[scale=0.04]{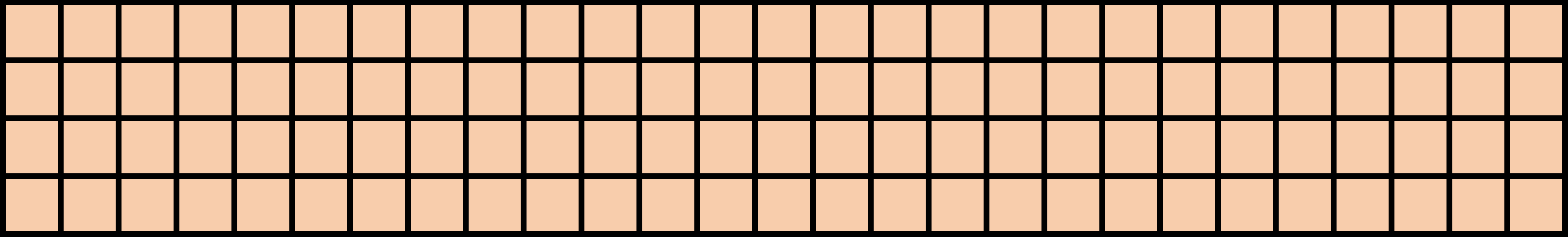}
    \end{subfigure}
    \begin{subfigure}[c][2.5cm][c]{0.3\linewidth}
        \centering
        \includegraphics[scale=0.04]{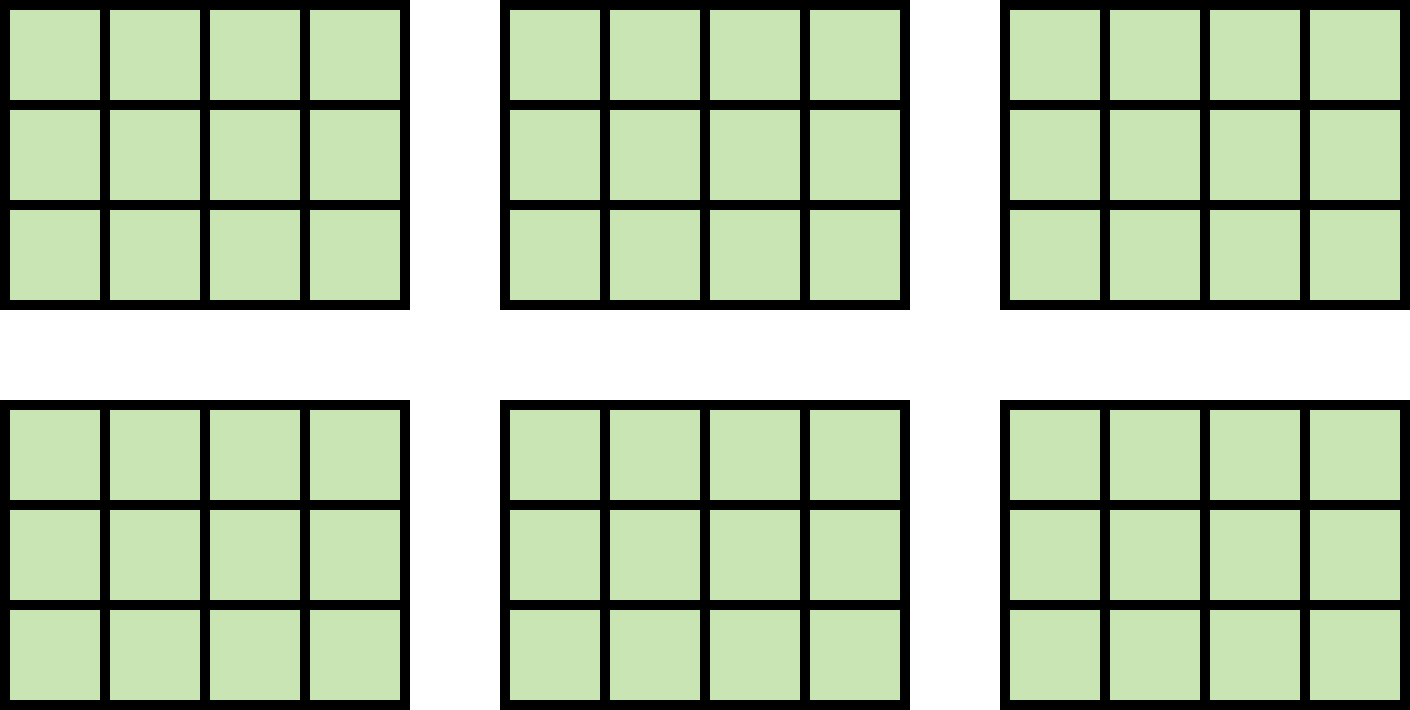}
    \end{subfigure}

    \begin{subfigure}[c][2.5cm][c]{0.3\linewidth}
        \centering
        \includegraphics[scale=0.04]{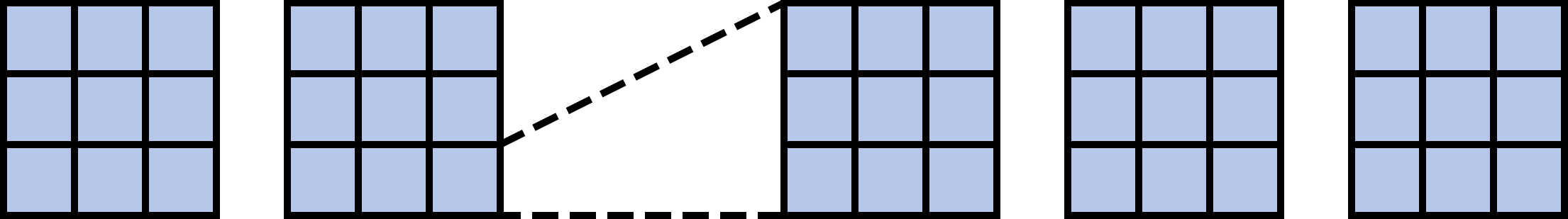}
    \end{subfigure}
    \begin{subfigure}[c][2.5cm][c]{0.3\linewidth}
        \centering
        \includegraphics[scale=0.04]{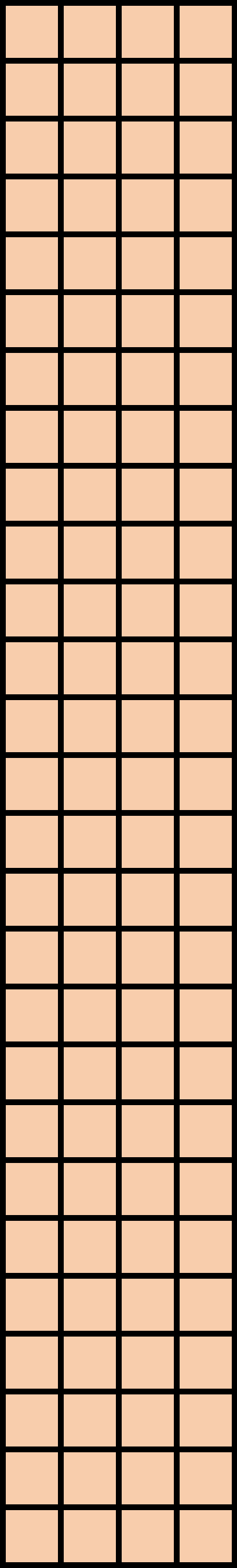}
    \end{subfigure}
    \begin{subfigure}[c][2.5cm][c]{0.3\linewidth}
        \centering
        \includegraphics[scale=0.04]{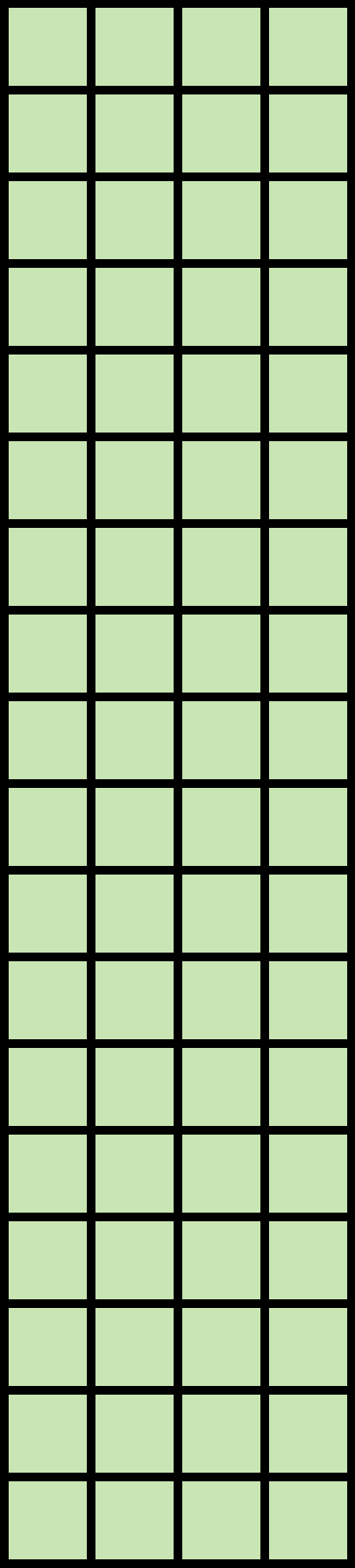}
    \end{subfigure}

    \begin{subfigure}[c][2.5cm][c]{0.3\linewidth}
        \centering
        \includegraphics[scale=0.04]{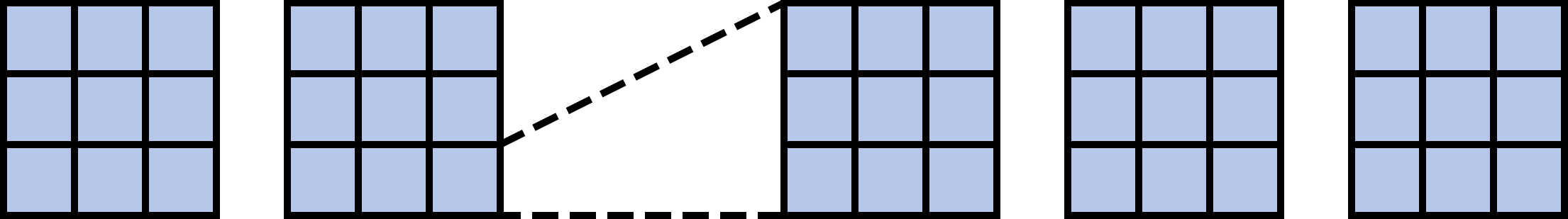}
    \end{subfigure}
    \begin{subfigure}[c][2.5cm][c]{0.3\linewidth}
        \centering
        \hfill
    \end{subfigure}
    \begin{subfigure}[c][2.5cm][c]{0.3\linewidth}
        \centering
        \hfill
    \end{subfigure}

    \begin{subfigure}[c][2.5cm][c]{0.3\linewidth}
        \centering
        \includegraphics[scale=0.04]{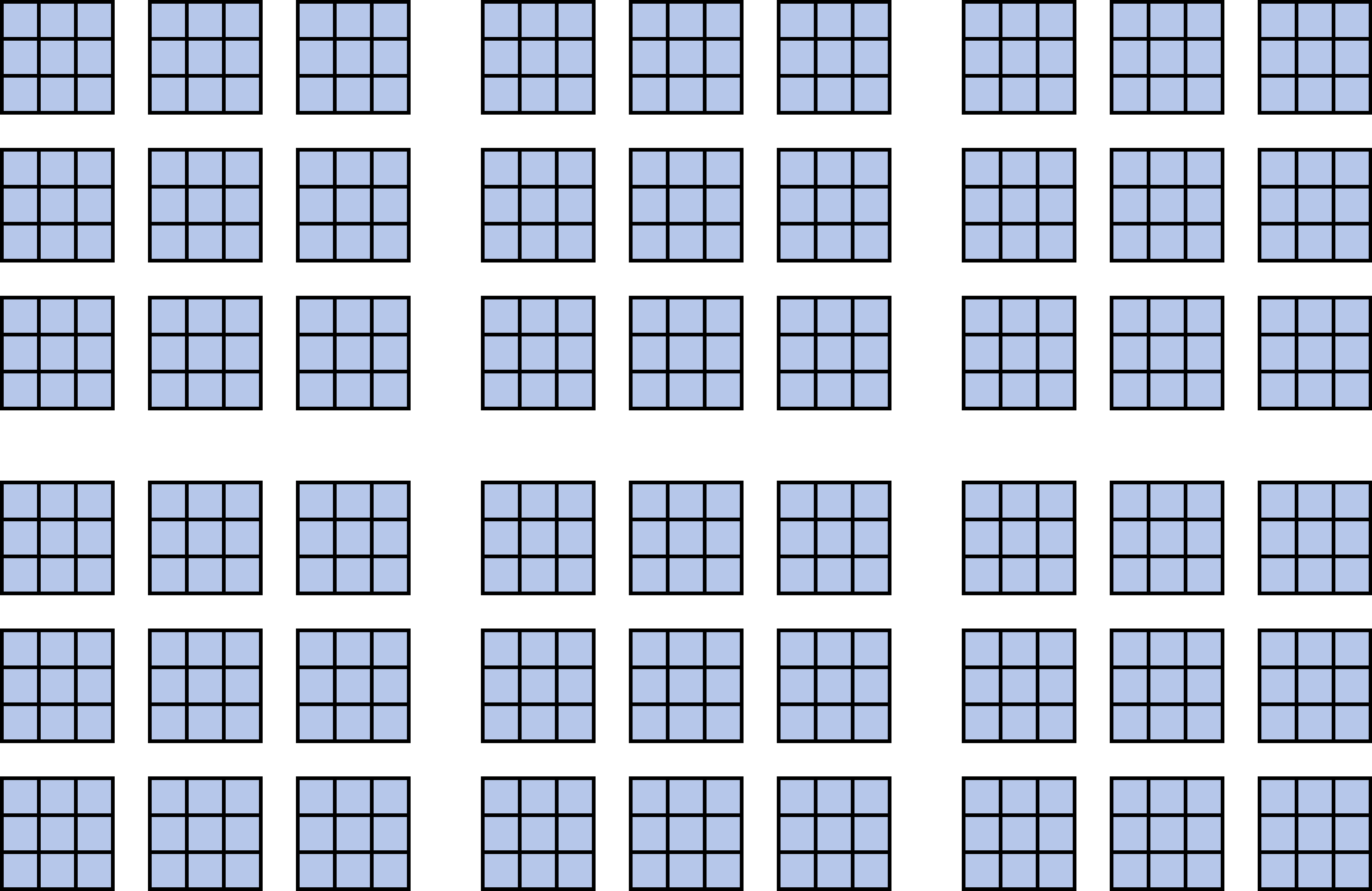}
    \end{subfigure}
    \begin{subfigure}[c][2.5cm][c]{0.3\linewidth}
        \centering
        \hfill
    \end{subfigure}
    \begin{subfigure}[c][2.5cm][c]{0.3\linewidth}
        \centering
        \hfill
    \end{subfigure}

    \begin{subfigure}[c][2.5cm][c]{0.3\linewidth}
        \centering
        \includegraphics[scale=0.04]{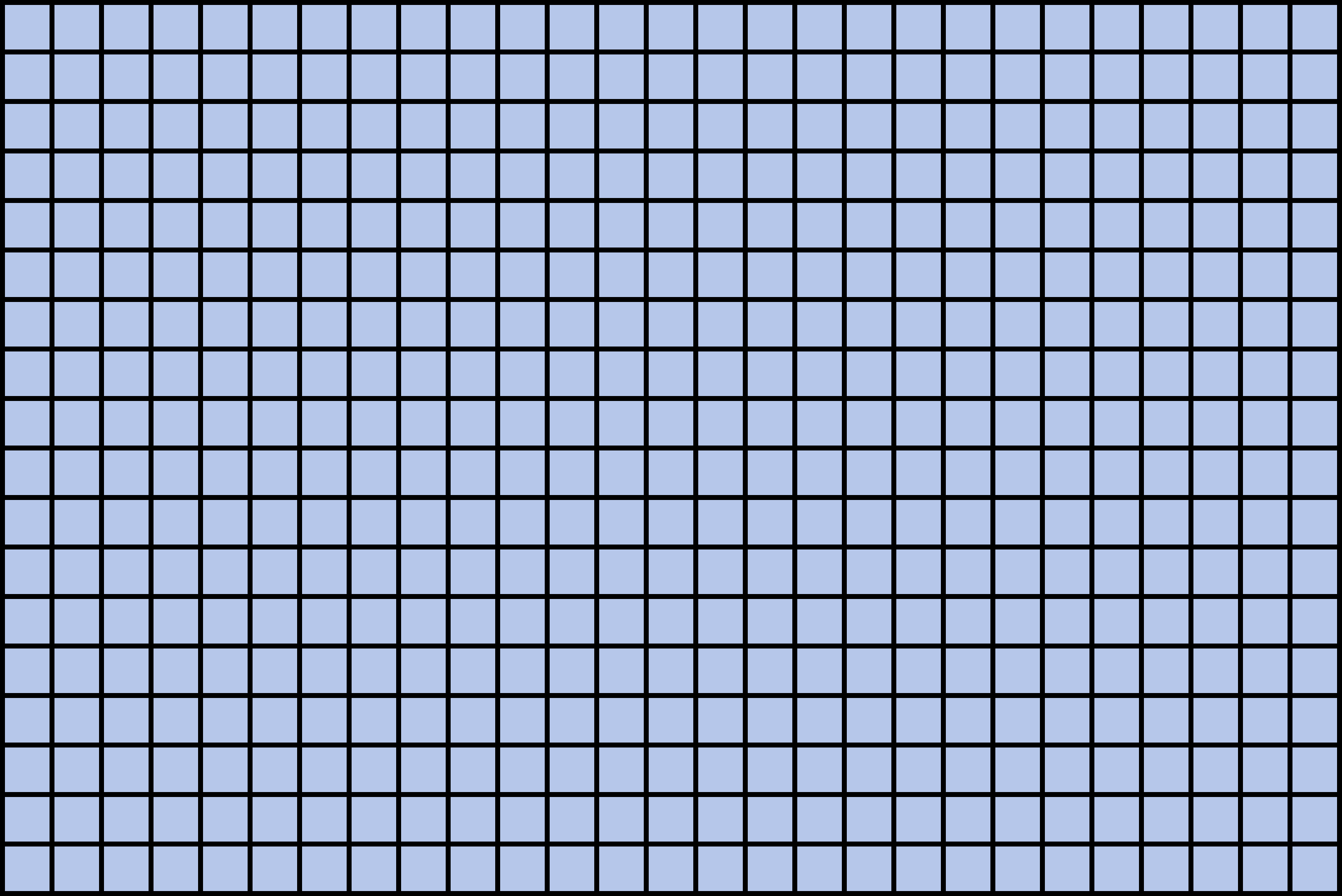}
    \end{subfigure}
    \begin{subfigure}[c][2.5cm][c]{0.3\linewidth}
        \centering
        \hfill
    \end{subfigure}
    \begin{subfigure}[c][2.5cm][c]{0.3\linewidth}
        \centering
        \hfill
    \end{subfigure}
\caption{A visualization of the 2D convolution arrangement.}
\label{fig:2d_convolution_arrangement}
\end{figure}

Let's denote the shape of the input as $(N, C, H, W)$, the shape of the filter as $(K, C, R, S)$, and the shape of the output as $(N, K, P, Q)$. In this code, we first tile the \texttt{input} tensor with the shape \texttt{(1, *filter.shape[1:])}, setting the \texttt{strides} to \texttt{(-1, -1, 1, 1)}. The last two dimensions are assigned a stride of \texttt{1}, reflecting the default stride of a basic 2D convolution. The first two dimensions are set to \texttt{-1}, indicating that we use the default strides, which are the same as the tile sizes, as the strides in a 2D convolution do not affect the $N$ and $C$ dimensions. After tiling, we obtain a hierarchical tensor with two levels: the first level has a shape of $(N, 1, P, Q)$, and the second level has a shape of $(1, C, R, S)$. Since the second dimension of the first level and the first dimension of the second level are singleton dimensions, we can squeeze them, resulting in a tensor that has a first level with a shape of $(N, P, Q)$ and a second level with a shape of $(C, R, S)$. Next, we ravel these two levels of this tensor, which results in a single-level tensor with a shape of $(N, P, Q, C, R, S)$. We then flatten the first three dimensions and the last three dimensions separately, yielding a tensor with a shape of $(N \cdot P \cdot Q, C \cdot R \cdot S)$. The meta-operations for the filter and the output tensors follow a similar process. The filter tensor is flattened and permuted, resulting in a tensor with a shape of $(C \cdot R \cdot S, K)$, while the output tensor is permuted and flattened to obtain a tensor that has a shape of $(N \cdot P \cdot Q, K)$. At this point, the convolution has been successfully mapped to a matrix multiplication based on the implicit GEMM algorithm. A visualization of the arrangement is provided in Figure~\ref{fig:2d_convolution_arrangement}.

\section{Evaluation}
\label{sec:evaluation}

To evaluate (1) whether NineToothed effectively simplifies Triton and (2) whether it maintains performance comparable to that of Triton, we conduct experiments from both code and performance perspectives. The code analysis includes raw metrics, cyclomatic complexity, Halstead metrics, and maintainability index, while the performance evaluation is carried out based on the execution of single kernel tasks and end-to-end model inference tasks.

\subsection{Experimental Setup}

The experiments are conducted on an NVIDIA A100 80GB PCIe GPU, with Triton version 3.0.0 and PyTorch version 2.4.1, and the following compute kernels are tested:

\begin{enumerate}
    \item \texttt{add}
    \item \texttt{addmm}
    \item \texttt{bmm}
    \item \texttt{conv2d}
    \item \texttt{mm}
    \item \texttt{rms\_norm}
    \item \texttt{rotary\_position\_embedding} (\texttt{rope})
    \item \texttt{scaled\_dot\_product\_attention} (\texttt{sdpa})
    \item \texttt{silu}
    \item \texttt{softmax}
\end{enumerate}

\noindent
Both NineToothed and Triton are used to implement these compute kernels, and the same algorithm is used for each kernel, ensuring, as much as possible, that any differences observed are not due to algorithmic discrepancies. For example, both the NineToothed and the Triton \texttt{conv2d} kernels follow the implicit GEMM algorithm, and both the NineToothed and the Triton \texttt{sdpa} kernels implement the FlashAttention-2~\cite{dao2023flashattention2} algorithm.

\subsection{Code Evaluation}

The code evaluation results are shown in Table~\ref{tab:code_comparison}.

\begin{table}[h]
\centering
\caption{Code evaluation results. The best results are \colorbox{green!20}{highlighted}.}
\begin{tabular}{llrrrrrrrrr}
\toprule
 &  & LOC & LLOC & SLOC & $G$ & $\eta$ & $N$ & $V$ & $D$ & $MI$ \\
\midrule
\multirow[c]{2}{*}{\texttt{add}} & Triton & {\cellcolor{green!20}} 17 & {\cellcolor{green!20}} 12 & {\cellcolor{green!20}} 12 & {\cellcolor{green!20}} 1 & 14 & 21 & 79.95 & 1.91 & 63.00 \\
 & NineToothed & 21 & {\cellcolor{green!20}} 12 & {\cellcolor{green!20}} 12 & 2 & {\cellcolor{green!20}} 3 & {\cellcolor{green!20}} 3 & {\cellcolor{green!20}} 4.75 & {\cellcolor{green!20}} 0.50 & {\cellcolor{green!20}} 87.91 \\
\multirow[c]{2}{*}{\texttt{addmm}} & Triton & 101 & 43 & 86 & {\cellcolor{green!20}} 2 & 89 & 153 & 990.79 & 4.35 & 43.12 \\
 & NineToothed & {\cellcolor{green!20}} 22 & {\cellcolor{green!20}} 12 & {\cellcolor{green!20}} 12 & {\cellcolor{green!20}} 2 & {\cellcolor{green!20}} 8 & {\cellcolor{green!20}} 9 & {\cellcolor{green!20}} 27.00 & {\cellcolor{green!20}} 1.00 & {\cellcolor{green!20}} 66.30 \\
\multirow[c]{2}{*}{\texttt{bmm}} & Triton & 93 & 44 & 82 & 2 & 87 & 159 & 1024.43 & 4.64 & 42.80 \\
 & NineToothed & {\cellcolor{green!20}} 39 & {\cellcolor{green!20}} 22 & {\cellcolor{green!20}} 29 & {\cellcolor{green!20}} 1 & {\cellcolor{green!20}} 3 & {\cellcolor{green!20}} 16 & {\cellcolor{green!20}} 25.36 & {\cellcolor{green!20}} 2.00 & {\cellcolor{green!20}} 60.75 \\
\multirow[c]{2}{*}{\texttt{conv2d}} & Triton & 129 & 56 & 110 & 2 & 116 & 237 & 1625.34 & 5.07 & 39.11 \\
 & NineToothed & {\cellcolor{green!20}} 25 & {\cellcolor{green!20}} 18 & {\cellcolor{green!20}} 16 & {\cellcolor{green!20}} 1 & {\cellcolor{green!20}} 2 & {\cellcolor{green!20}} 4 & {\cellcolor{green!20}} 4.00 & {\cellcolor{green!20}} 1.00 & {\cellcolor{green!20}} 68.13 \\
\multirow[c]{2}{*}{\texttt{mm}} & Triton & 85 & 39 & 74 & {\cellcolor{green!20}} 2 & 75 & 132 & 822.20 & 4.53 & 44.61 \\
 & NineToothed & {\cellcolor{green!20}} 44 & {\cellcolor{green!20}} 24 & {\cellcolor{green!20}} 31 & 3 & {\cellcolor{green!20}} 5 & {\cellcolor{green!20}} 11 & {\cellcolor{green!20}} 25.54 & {\cellcolor{green!20}} 2.00 & {\cellcolor{green!20}} 59.77 \\
\multirow[c]{2}{*}{\texttt{rms\_norm}} & Triton & 27 & 12 & 20 & {\cellcolor{green!20}} 1 & 20 & 33 & 142.62 & 2.75 & 61.24 \\
 & NineToothed & {\cellcolor{green!20}} 21 & {\cellcolor{green!20}} 11 & {\cellcolor{green!20}} 13 & 2 & {\cellcolor{green!20}} 11 & {\cellcolor{green!20}} 14 & {\cellcolor{green!20}} 48.43 & {\cellcolor{green!20}} 2.57 & {\cellcolor{green!20}} 81.24 \\
\multirow[c]{2}{*}{\texttt{rope}} & Triton & {\cellcolor{green!20}} 53 & {\cellcolor{green!20}} 26 & 43 & {\cellcolor{green!20}} 2 & 52 & 114 & 649.85 & {\cellcolor{green!20}} 4.04 & 49.17 \\
 & NineToothed & 56 & 34 & {\cellcolor{green!20}} 39 & 5 & {\cellcolor{green!20}} 16 & {\cellcolor{green!20}} 29 & {\cellcolor{green!20}} 116.00 & 4.09 & {\cellcolor{green!20}} 51.60 \\
\multirow[c]{2}{*}{\texttt{sdpa}} & Triton & 120 & {\cellcolor{green!20}} 43 & 106 & {\cellcolor{green!20}} 2 & 56 & 87 & 505.24 & {\cellcolor{green!20}} 2.84 & 45.17 \\
 & NineToothed & {\cellcolor{green!20}} 60 & 44 & {\cellcolor{green!20}} 42 & 3 & {\cellcolor{green!20}} 30 & {\cellcolor{green!20}} 58 & {\cellcolor{green!20}} 284.60 & 4.25 & {\cellcolor{green!20}} 55.75 \\
\multirow[c]{2}{*}{\texttt{silu}} & Triton & {\cellcolor{green!20}} 16 & {\cellcolor{green!20}} 11 & {\cellcolor{green!20}} 11 & {\cellcolor{green!20}} 1 & 13 & 18 & 66.61 & 1.80 & 64.38 \\
 & NineToothed & 19 & {\cellcolor{green!20}} 11 & {\cellcolor{green!20}} 11 & 2 & {\cellcolor{green!20}} 3 & {\cellcolor{green!20}} 3 & {\cellcolor{green!20}} 4.75 & {\cellcolor{green!20}} 0.50 & {\cellcolor{green!20}} 89.51 \\
\multirow[c]{2}{*}{\texttt{softmax}} & Triton & 31 & 17 & 24 & {\cellcolor{green!20}} 1 & 20 & 27 & 116.69 & 3.00 & 58.55 \\
 & NineToothed & {\cellcolor{green!20}} 24 & {\cellcolor{green!20}} 14 & {\cellcolor{green!20}} 14 & 2 & {\cellcolor{green!20}} 6 & {\cellcolor{green!20}} 6 & {\cellcolor{green!20}} 15.51 & {\cellcolor{green!20}} 1.00 & {\cellcolor{green!20}} 81.90 \\
\bottomrule
\end{tabular}
\label{tab:code_comparison}
\end{table}

\subsubsection{Raw Metrics}

Raw metrics (LOC, LLOC, SLOC) provide a direct measure of code volume. In nearly all kernels, NineToothed implementations are substantially shorter and more concise compared to their Triton counterparts. Since NineToothed and Triton employ different auto-tuning mechanisms, the LOC and SLOC values presented here may not be convincing. Nevertheless, the LLOC follow the same trend in most cases, indicating that NineToothed avoids boilerplate and offers more expressive syntax. The only exceptions occur in very simple kernels like \texttt{add} and \texttt{silu}, where reductions are marginal or even slightly negative, showing that NineToothed may introduce minor overhead in extremely short functions. However, the general trend clearly supports NineToothed’s effectiveness at reducing code size.

\subsubsection{Cyclomatic Complexity}

Cyclomatic complexity ($G$) measures the number of independent paths through a program. Lower values generally indicate easier-to-understand code. NineToothed's complexity matches Triton’s complexity. There are a few cases where complexity increases, suggesting that in certain cases the abstraction introduced by NineToothed may require slightly more intricate control flow. However, this tradeoff is both rare and modest, as neither NineToothed nor Triton kernels are theoretically cyclomatically complex. Overall, NineToothed maintains cyclomatic simplicity in the majority of the tested kernels.

\subsubsection{Halstead Metrics}

Halstead metrics, which include measures like program vocabulary ($\eta$), program length ($N$), volume ($V$), and difficulty ($D$), provide insight into cognitive load. NineToothed shows consistently lower values across these metrics, reflecting reduced mental effort required to comprehend the code. The vocabulary and volume also decrease significantly in all complex kernels. Statistical analysis shows that the Halstead code volume of compute kernels implemented with NineToothed ranges from only 0.25\% to 56.33\% of that written using Triton. These drastic reductions across all Halstead measures reinforce the conclusion that NineToothed greatly simplifies Triton.

\subsubsection{Maintainability Index}

The maintainability index ($MI$) is a composite metric that incorporates the Halstead volume, the cyclomatic complexity, and the source lines of code to assess code maintainability. A higher index indicates easier maintenance. Based on the prior analyses, NineToothed consistently reduces the SLOC and the Halstead volume, all of which directly contribute to a higher maintainability index. NineToothed shows higher values in all tested kernels, reflecting that NineToothed exhibits greater maintainability compared to Triton.

\subsection{Performance Evaluation}

To evaluate the performance of NineToothed, we conduct experiments on both single compute kernel tasks and end-to-end model inference tasks. In these experiments, we compare the performance of NineToothed with that of Triton. PyTorch is also included, but it should be viewed only as a supplementary point of reference, since its operator implementations may differ in terms of algorithms from the other two.

\subsubsection{Single Compute Kernel Tasks}

The following tasks are involved, where each task is represented in the form of \texttt{op(*args, **kwargs)}, with \texttt{op} denoting the name of the operator and the values of \texttt{args} and \texttt{kwargs} corresponding to the shapes of the respective argument tensors:

\begin{enumerate}
    \item \texttt{add((16777216,), (16777216,))}
    \item \texttt{addmm((4096, 4096), (4096, 4096), (4096, 4096), beta=(), alpha=())}
    \item \texttt{bmm((4, 2048, 2048), (4, 2048, 2048))}
    \item \texttt{conv2d((4, 512, 14, 14), (512, 512, 3, 3))}
    \item \texttt{mm((4096, 4096), (4096, 4096))}
    \item \texttt{rms\_norm((4096, 4096))}
    \item \texttt{rope((4, 1024, 48, 64), (1024, 32), (1024, 32))}
    \item \texttt{sdpa((4, 48, 1024, 64), (4, 48, 1024, 64), (4, 48, 1024, 64))}
    \item \texttt{silu((16777216,))}
    \item \texttt{softmax((4096, 4096))}
\end{enumerate}

\noindent
Each tensor has a data type of \texttt{float16}.

\begin{figure}[h]
\centering
\includegraphics[width=0.9\linewidth]{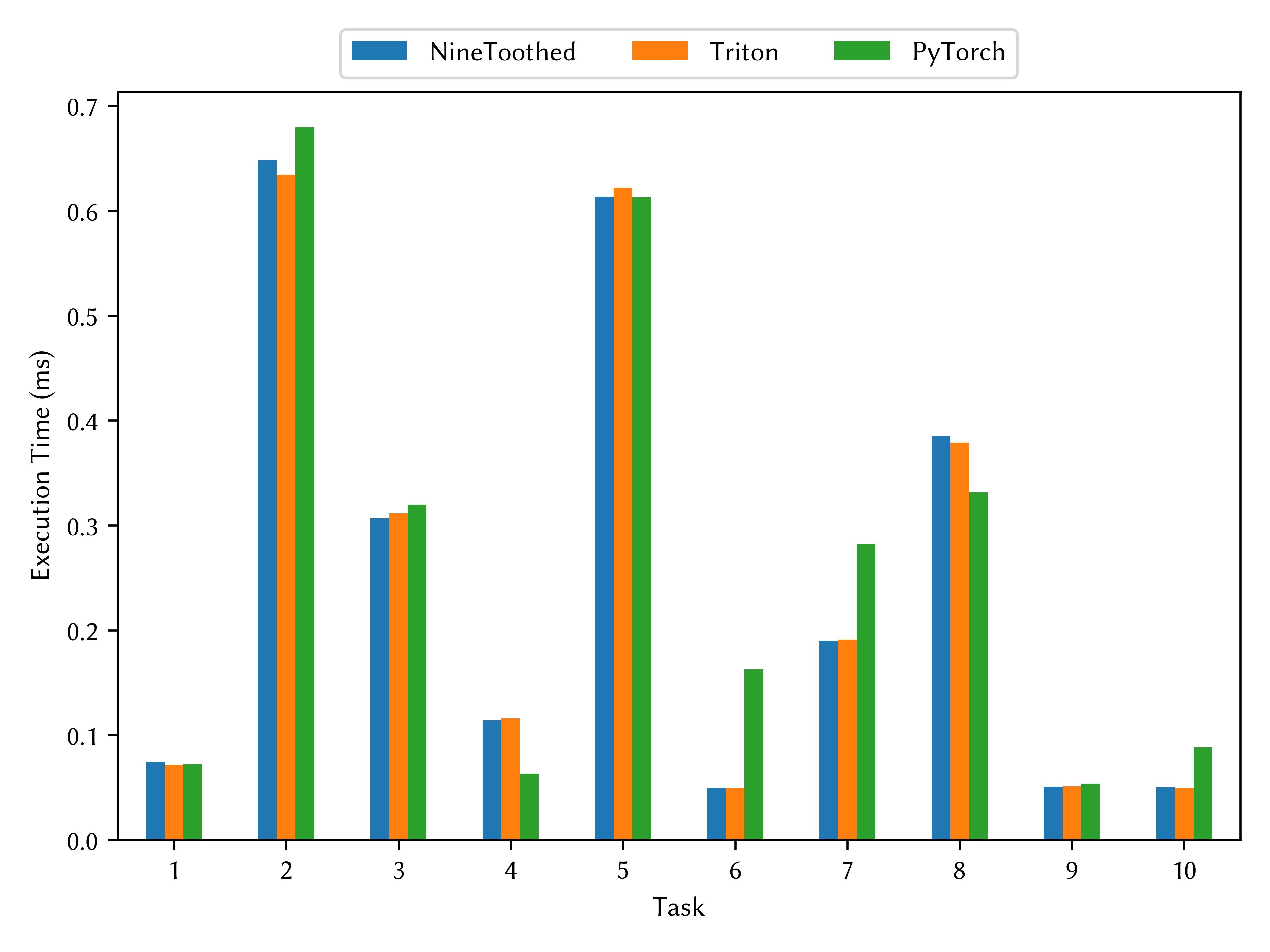}
\caption{Performance of single compute kernel tasks.}
\label{fig:microbenchmark_results}
\end{figure}

As shown in Figure~\ref{fig:microbenchmark_results}, the performance of NineToothed and Triton closely aligns across all tasks. Statistical analysis reveals that the relative percentage difference between NineToothed and Triton ranges from a minimum of -1.58\% to a maximum of 3.93\%, with an average of 0.37\%. This indicates that NineToothed's performance in the single compute kernel tasks is comparable to that of Triton.

\subsubsection{End-to-End Model Inference Tasks}

The end-to-end model inference experiments are conducted using the \texttt{deepseek-ai/DeepSeek-R1-Distill-Llama-8B} model, with a batch size of 2 and an input length of 32 tokens. These experiments target tasks with three distinct output lengths: 128, 512, and 2048 tokens. To evaluate the performance impact of custom kernel implementations, the following components of the original PyTorch model are replaced with hand-written operators implemented using both NineToothed and Triton:

\begin{enumerate}
    \item \texttt{Attention}
    \item \texttt{Linear}
    \item \texttt{RMSNorm}
    \item \texttt{SiLU}
\end{enumerate}

\noindent
In particular, the \texttt{Attention} module incorporates custom \texttt{rope} kernels written in both DSLs. Performance is assessed based on model inference throughput, measured in tokens per second. Each task includes a single warmup iteration, followed by three measured iterations; the reported throughput corresponds to the average of these three measurements.

\begin{figure}[h]
\centering
\includegraphics[width=0.9\linewidth]{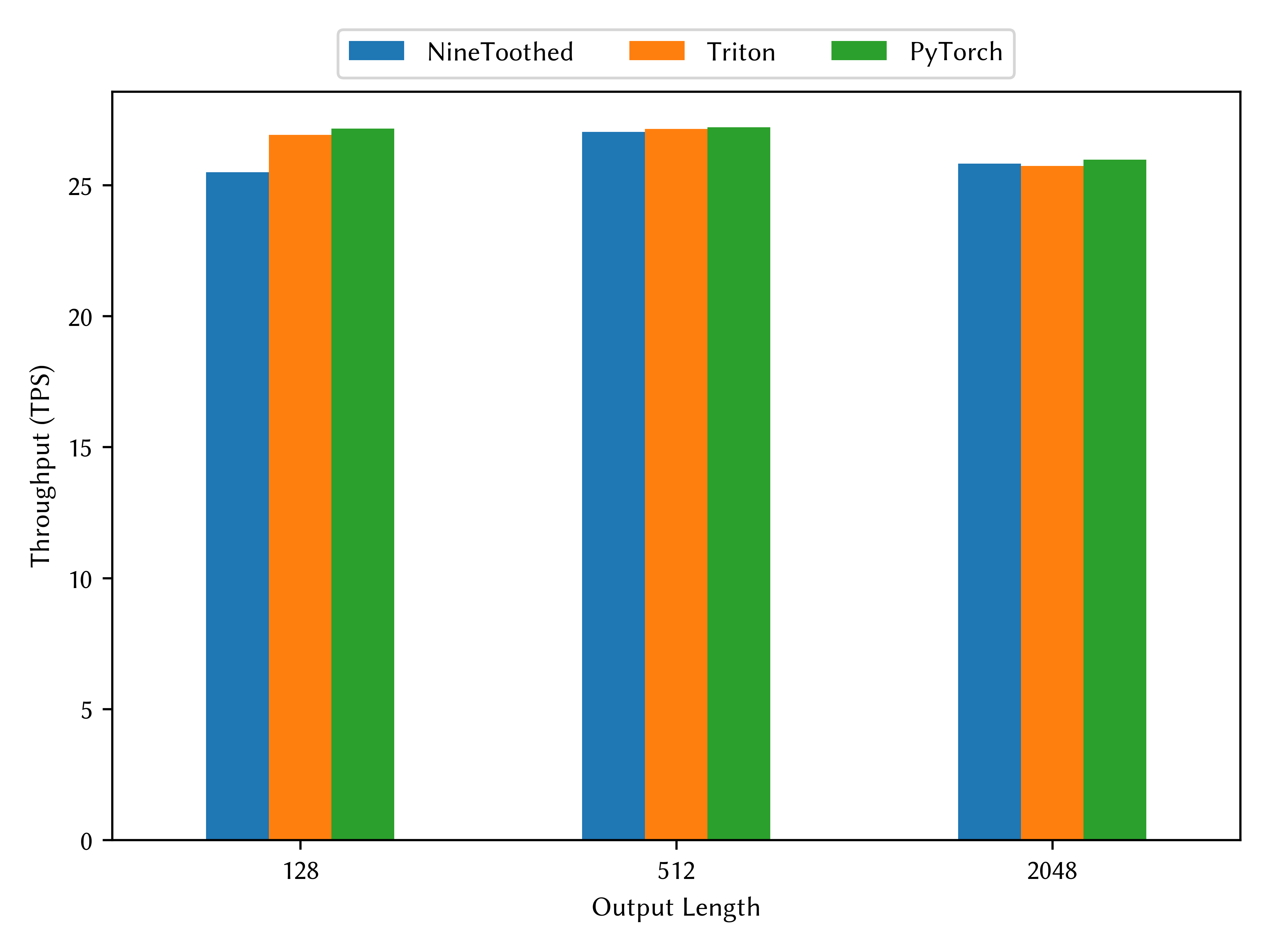}
\caption{Performance of end-to-end model inference tasks.}
\label{fig:benchmark_results}
\end{figure}

The experimental results are shown in Figure~\ref{fig:benchmark_results}. Statistical analysis indicates that the relative percentage difference between NineToothed and Triton ranges from a minimum of -5.32\% to a maximum of 0.33\%, with an average of -1.79\%. These findings suggest that NineToothed's performance in the end-to-end model inference tasks is also comparable to that of Triton.

\section{Related Work}
\label{sec:related_work}

Triton is a domain-specific language (DSL) and compiler designed to facilitate the development of high-performance compute kernels on GPUs, abstracting away much of the complexity inherent in GPU programming. Its primary objective is to narrow the gap between high-level programming languages and low-level, hand-written GPU code, thereby making GPU programming more accessible to users without specialized expertise. Although Triton abstracts away many low-level details, it retains a typical parallel programming model. In contrast, NineToothed extends this abstraction further, allowing developers to express tiled computations by writing serial code.

Graphene, an intermediate representation (IR), is dedicated to optimizing tensor computations at a lower level. Unlike existing tensor IRs, which often struggle to represent the intricate data-to-thread mappings essential for GPU tensor instructions, Graphene introduces hierarchical tensors as first-class objects. These tensors can be decomposed into tiles, enabling efficient data-to-thread mappings and optimization of tensor computations. NineToothed is heavily influenced by Graphene, incorporating concepts such as hierarchical tensors. However, while Graphene serves as a low-level IR, NineToothed is a high-level DSL designed for developers to use directly.

Numba is a just-in-time (JIT) compiler that can enhance the performance of numerical or array-oriented Python code by parallelizing the execution and leveraging LLVM to generate optimized machine code~\cite{10.1145/2833157.2833162}. It is designed to work seamlessly with NumPy and can be used on NVIDIA GPUs and a variety of CPUs. However, unlike NineToothed, Numba is designed to be more versatile regarding use cases, focusing on general numerical computing rather than deep learning tasks. As a result, Numba does not specifically cater to popular deep-learning frameworks like PyTorch. When writing CUDA compute kernels, Numba maps CUDA programming from C++ to Python directly, meaning that users are still expected to have a concrete understanding of low-level details. NineToothed, on the other hand, abstracts away much of the complexity of CUDA programming.


\section{Conclusion}
\label{sec:conclusion}

In this paper, we proposed NineToothed, a DSL for expressing tiled computations with serial programming. By introducing tensor-oriented metaprogramming, NineToothed abstracts away low-level details such as pointer arithmetic and memory access. By designing the arrange-and-apply paradigm, NineToothed is able to convert serial code into high-performance parallel code. Our results demonstrate that NineToothed achieves performance comparable to that of Triton with significantly less code while offering superior readability and maintainability.



\section*{Availability}

NineToothed is publicly available at \href{https://github.com/InfiniTensor/ninetoothed}{https://github.com/InfiniTensor/ninetoothed}.

\bibliographystyle{plain}
\bibliography{refs}

\end{document}